\documentclass[pdflatex,sn-basic]{sn-jnl}

\usepackage{graphicx}
\usepackage{multirow}
\usepackage{amsmath,amssymb,amsfonts}
\usepackage{xcolor}
\usepackage{textcomp}
\usepackage{booktabs}
\usepackage{subcaption}

\begin{document}

\title[Smartphone turbidity measurement and device transfer]
{Controlled-illumination smartphone imaging for quantitative turbidity measurement and cross-device transfer}

\author*[1]{\fnm{Sebasti\'an} \sur{Olivares}}\email{sebastian.olivares@uss.cl}
\author[3,5]{\fnm{Addles} \sur{Castillo}}\email{a.castilloespinosa@uandresbello.edu}
\author*[2]{\fnm{Patricia} \sur{Vel\'asquez}}\email{patricia.velasquez@unab.cl}
\author[2,4]{\fnm{Roxana} \sur{Arce}}\email{roxana.arce@unab.cl}
\author[3]{\fnm{Nicol\'as} \sur{Richardson}}\email{n.richardsonrojas@uandresbello.edu}
 
\affil[1]{\orgdiv{Facultad de Ingenier\'ia}, \orgname{Universidad San Sebasti\'an},
\orgaddress{\street{Bellavista 7}, \city{Santiago}, \postcode{8420524},
\state{Metropolitana}, \country{Chile}}}

\affil[2]{\orgdiv{Departamento de Ciencias Qu\'imicas, Facultad de Ciencias Exactas},
\orgname{Universidad Andr\'es Bello}, \orgaddress{\street{Avenida Rep\'ublica 275},
\city{Santiago}, \postcode{8370146}, \state{Metropolitana}, \country{Chile}}}

\affil[3]{\orgdiv{Departamento de F\'isica y Astronom\'ia, Facultad de Ciencias Exactas},
\orgname{Universidad Andr\'es Bello}, \orgaddress{\street{Avenida Rep\'ublica 220},
\city{Santiago}, \postcode{8370146}, \state{Metropolitana}, \country{Chile}}}

\affil[4]{\orgname{Millennium Institute on Green Ammonia as Energy Vector (MIGA)}, \orgaddress{\street{Av. Vicuña Mackenna 4860},
\city{Santiago}, \postcode{7820436}, \state{Metropolitana}, \country{Chile}}}

\affil[5]{\orgname{Millennium Institute for Subatomic Physics at High Energy Frontier (SAPHIR)}, \orgaddress{\street{Fernández Concha 700},
\city{Santiago}, \postcode{7591538}, \state{Metropolitana}, \country{Chile}}}

\abstract{%
Smartphone imaging could provide a low-cost turbidity measurement method if
illumination and acquisition geometry are controlled. We evaluated a
3D-printed enclosure with fixed illumination and a submerged optical target
using 162 milk-based suspensions spanning 0.16--50.2 nephelometric turbidity units (NTU), with a
portable turbidimeter as the reference. An EfficientNetB0 regression model evaluated using
repeated five-fold grouped cross-validation achieved a mean absolute error
(MAE) of 1.208 NTU and $R^2=0.9816$, indicating that predictions followed the
reference measurements closely. The MAE was 0.335 NTU below 1 NTU and 0.449 NTU
from 1 to below 5 NTU; classification accuracy at the operational 5-NTU
threshold was 0.9753.
When applied without retraining to 103 paired images from a second smartphone,
prediction error remained comparable to that on the development smartphone
below 10 NTU, but turbidity was increasingly underestimated at higher values. The results demonstrate the measurement performance attainable
with a controlled smartphone acquisition system, quantify an important
device-dependent transfer limitation, and motivate standardized multi-device
acquisition and prospective natural-water validation.%
}


\maketitle

% ============================================================================
\section{Introduction}\label{sec:introduction}

\subsection{Motivation and operational context}

Turbidity describes the reduction in water clarity caused by suspended and
colloidal material that scatters and absorbs light. It is monitored not only
in drinking-water production and post-treatment quality control, but also in
surface-water assessment, environmental monitoring, aquaculture, and
agricultural water management (\citealp{EPA_TurbidityFactsheet};
\citealp{FAO_Aquaculture}; \citealp{FAO_Irrigation}). Changes in turbidity can indicate variation in source-water
conditions, erosion and runoff, sediment resuspension, algal growth, treatment
performance, distribution-system disturbances, or contamination events.
Suspended material can also reduce light penetration and affect aquatic
processes, interfere with disinfection, and alter the appearance and
acceptability of drinking water. Turbidity should nevertheless be interpreted
together with microbiological, chemical, hydrological, and other measurements
because neither high nor low turbidity alone establishes overall water safety
or quality \citep{WHO_Turbidity_2017,EPA_TurbidityFactsheet}.

ISO 7027-1:2016 describes two quantitative optical principles for determining
turbidity in water \citep{ISO7027_1_2016}. Nephelometry measures radiation
scattered by suspended particles and is particularly applicable to
low-turbidity waters, including drinking water, whereas turbidimetry measures
attenuation of the incident radiant flux and is more applicable to highly
turbid waters. Turbidity reporting units depend on the optical method and
instrument configuration. Formazin nephelometric units (FNU) are associated
with infrared nephelometric measurements, whereas NTU is commonly used for
EPA-method measurements; ISO attenuation measurements are expressed in
formazin attenuation units (FAU). Measurements from different instrument
designs are not necessarily interchangeable because their responses can
differ with sample color, particle size, and particle concentration
\citep{USGS_TurbidityReporting2004}. In the present study, reference values
were obtained using a ratio-nephelometric instrument and are reported in NTU,
as provided by that instrument; no conversion to FNU was performed.

Extending instrument-based monitoring to more locations or increasing
sampling frequency can be constrained by equipment cost, calibration and maintenance
requirements, sample handling, and the availability of trained operators. A
compact system combining controlled illumination, fixed acquisition geometry,
and smartphone imaging could therefore complement conventional measurements
by increasing spatial or temporal screening coverage and identifying samples
that warrant confirmatory analysis. Such a system should be positioned as a
screening and decision-support tool rather than as a replacement for validated
laboratory or regulatory procedures.

Turbidity requirements depend on water source, treatment process, and
regulatory context. For example, United States surface-water treatment rules
specify limits as low as 0.3 NTU for at least 95\% of monthly measurements
and a maximum of 1 NTU for conventional or direct filtration, while some
other filtration categories are subject to limits that cannot exceed 5 NTU
\citep{EPA_NPDWR,EPA_SWTR}. WHO guidance likewise emphasizes maintaining low
turbidity to support effective disinfection \citep{WHO_Turbidity_2017}. In
this study, 5 NTU is used as a predefined operational threshold for evaluating
screening decisions. It should not be interpreted as a universal potability
boundary or as the preferred target for all treated-water systems.

\subsection{Smartphone imaging and the device-dependence problem}

Smartphones provide high-resolution cameras, substantial onboard computation,
and broad availability, making them attractive platforms for portable optical
measurements. When a model is trained against reference NTU measurements and
the optical conditions are constrained, a smartphone image can provide a
rapid quantitative turbidity estimate. This creates a practical path toward
low-cost, high-frequency screening and eventually toward an automated mobile
application. Previous smartphone turbidity systems have used controlled
illumination, optical fibers, external lenses, proximity sensors, reference
targets, or structured acquisition procedures to relate recorded optical
signals to reference turbidity \citep{Koydemir2019,Toivanen2013,Hussain2016}.
Other approaches have attempted to infer turbidity more directly from camera
images or above-water reflectance measurements
\citep{LeeuwBoss2018,Wilches2022BMVC,Jantarakasem2024AQUA}.

The accessibility of smartphone imaging is accompanied by an important
reproducibility problem. Recorded images depend not only on the water sample
but also on illumination, viewing geometry, exposure, white balance, optics,
sensor spectral response, compression, and manufacturer-specific image
processing. Camera selection has been shown to affect the capacity to predict
turbidity from mobile-device images \citep{Villamil2021}. Physical control of
illumination and geometry can suppress several acquisition variables, but it
does not necessarily make the image response of different smartphones
equivalent.

The resulting device-domain shift is particularly important for models
trained on one phone and deployed on another. Strong in-domain performance
may reflect device-specific correlations that do not transfer, and a
calibration that improves full-range numerical error may not preserve
performance around an operational decision threshold. Credible evaluation
therefore requires separation of internal model validation, direct
cross-device performance without retraining (commonly called \emph{zero-shot}
transfer), and explicitly post-hoc calibration. Large-scale deployment would
require acquisition and calibration procedures that standardize the relevant
responses of different smartphone cameras.

\subsection{Study objectives and contributions}

This study combines a primary continuous-NTU regression analysis with
secondary classification, repeatability, and paired-device experiments. It
addresses four questions:

\begin{enumerate}
  \item Can controlled-illumination images from the development smartphone
  predict continuous turbidity under repeated grouped cross-validation?

  \item What does a secondary spatial-representation classification experiment
  reveal about the image regions used by the model, and does
  representation affect transfer to another smartphone?

  \item How repeatable is the combined refill, reference-measurement,
  settling, imaging, and prediction procedure under one tested low-turbidity
  condition?

  \item How does a regression model trained on the development smartphone
  perform without retraining on paired images from a second smartphone, and
  what can explicitly post-hoc calibration recover?
\end{enumerate}

The principal contributions are: (i) continuous turbidity regression using
NTU ground truth for 162 physical samples and repeated grouped validation, in
which the samples were divided into five held-out subsets and every sample was
tested only by a model that had not been trained on it; (ii) range-specific evaluation below
10 NTU and around a predefined 5-NTU operational threshold; (iii) a
nine-fill repeatability analysis, with one additional observation excluded
because of an independently confirmed acquisition or labeling error and
retained for sensitivity analysis; (iv) direct transfer without
retraining to paired images from 103 physical samples; and (v) secondary analyses
of spatial representation, input orientation, and cross-fitted post-hoc
device calibration. Cross-fitting means that the correction applied to each
sample was learned only from other samples, preventing that sample's own
reference value from entering its correction. The primary claims concern
internal grouped validation on the development smartphone and uncalibrated
transfer to the second smartphone. Calibration and orientation
normalization are interpreted as post-hoc sensitivity analyses.

% ============================================================================
\section{Related work}\label{sec:related_work}

\subsection{Controlled optical attachments and smartphone readers}

Several smartphone turbidity systems use dedicated opto-mechanical components
to create a more reproducible optical measurement. Hussain et al. used an
infrared source and a smartphone proximity sensor to measure scattered light
over a calibrated turbidity range \citep{Hussain2016}. Koydemir et al.
developed a 3D-printed smartphone attachment containing white light-emitting diodes (LEDs), optical
fibers, and an external lens, combining nephelometric and transmittance
measurements to cover a broad turbidity range \citep{Koydemir2019}. Toivanen
et al. combined a mobile phone with a small measurement device and evaluated
mobile-image analysis for water transparency and turbidity in a participatory
field study \citep{Toivanen2013}. Parra et al. combined a programmable RGB
light source and a light-dependent resistor in a low-cost optical sensor and
compared multiple machine-learning approaches for both continuous turbidity
quantification and classification across samples representing four turbidity
sources \citep{Parra2024}.

These studies demonstrate the value of controlling the light path and
phone-to-sample geometry. Their optical configurations differ from the
present enclosure, which images the attenuation and distortion of a submerged
high-contrast target under fixed internal illumination. Nevertheless, they
establish that relatively inexpensive smartphone-based systems can provide
quantitative or screening-level optical information when acquisition is
sufficiently constrained.

\subsection{Image-based and minimal-hardware approaches}

A complementary research direction uses the smartphone camera more directly.
The HydroColor application treats the camera as an above-water radiometer and
uses structured image acquisition to estimate remote-sensing reflectance and
turbidity in natural water bodies \citep{LeeuwBoss2018}. Wilches et al.
formulated turbidity estimation as a computer-vision problem using smartphone
images and investigated capture and modeling strategies for camera-based
estimation \citep{Wilches2022BMVC}. Jantarakasem et al. subsequently evaluated
smartphone images for drinking-water turbidity estimation using particulate
standards and discrete turbidity ranges \citep{Jantarakasem2024AQUA}.
Nie et al. compared customized convolutional neural network (CNN) architectures for classifying images of
ten calcium-carbonate turbidity levels acquired under fixed laboratory
conditions and also evaluated performance after synthetic image-noise
augmentation \citep{Nie2025}. Machine-vision methods have also been studied
for turbidity estimation from underwater imagery. Rudy and Wilson combined
YOLOv8 image classification with regression to estimate continuous turbidity
from laboratory-mesocosm and in-situ underwater images \citep{Rudy}.

These approaches differ from the present system. HydroColor measures
above-water reflectance; underwater methods analyze scene degradation; and
some camera-only methods use less constrained views. Here, the phone images a
submerged high-contrast target inside a fixed enclosure under controlled
internal illumination. These systems consequently measure different optical
phenomena and should not be compared only through headline accuracy values.
Above-water reflectance, side-scattered light, transmitted light, target
attenuation, and underwater image degradation have different sensitivities to
particle composition, color, illumination, distance, and camera processing.
Evaluation against a reference turbidimeter and clear reporting of the
acquisition geometry are therefore necessary for interpreting model
performance.

\subsection{Cross-device generalization and calibration}

Cross-device generalization remains a central obstacle for camera-based water
measurements. Villamil et al. directly examined mobile-camera selection and
showed that the capacity to predict turbidity depends on the device used to
acquire the images \citep{Villamil2021}. Differences may arise from sensor
response, lens characteristics, exposure and white-balance behavior, tone
mapping, sharpening, denoising, compression, or software updates.

Controlled enclosures can reduce variation in illumination and geometry, but
the resulting images still pass through device-specific camera pipelines.
Consequently, validation on the training device is insufficient to establish
deployment performance on another phone. Paired imaging of the same physical
samples provides a direct way to characterize device-dependent prediction
differences. Device calibration may reduce these differences, but calibration
must be evaluated without allowing a sample's own reference value to influence
the correction applied to that sample. Even cross-fitted calibration remains
exploratory when its form is chosen after observing the transfer behavior.

\subsection{Position of the present study}

In the available literature, controlled acquisition, machine-learning
prediction against instrument-derived reference values, sample-level
validation, repeatability assessment, and cross-device transfer are commonly
examined separately \citep{Parra2024,Nie2025,Rudy}. The present work brings
these elements together in one
acquisition and validation framework. The primary model predicts continuous
turbidity over 0.16--50.2 NTU from images acquired under controlled
illumination, with reference labels obtained from a calibrated
turbidimeter. Predictions are evaluated at the physical-sample level using
medians of repeated out-of-fold estimates rather than a single image-level
train--test split.

The study further evaluates low-NTU numerical performance and decision
behavior around a predefined 5-NTU operational threshold, short-term
procedural repeatability, and direct transfer to paired images from a second
smartphone. It also examines whether spatial representations that perform
similarly in a secondary Samsung classification experiment transfer similarly
between camera pipelines. Uncalibrated prediction on the second device is
treated as the primary device-transfer result; orientation normalization and
three cross-fitted calibration mappings are reported as secondary post-hoc
analyses. The contribution is the combined evaluation of continuous estimation,
low-range decision behavior, repeatability, and paired cross-device transfer
within a controlled acquisition system.

From a measurement perspective, the work characterizes the complete chain from
reference measurement and controlled optical acquisition to image-based NTU
estimation. Its principal methodological contribution is the combined,
sample-level evaluation of internal prediction error, short-term procedural
repeatability, and transfer of the fitted measurement model between two camera
pipelines. This design separates performance on the development device from
cross-device performance and reveals differences between device measurements that would be missed by
validation on a single device.

% ============================================================================
\section{Materials and methods}\label{sec:methods}

\subsection{Controlled imaging enclosure and hardware}\label{subsec:hardware}

\begin{figure}[t]
\centering

\begin{subfigure}[t]{0.48\textwidth}
\centering
\includegraphics[width=\linewidth]
{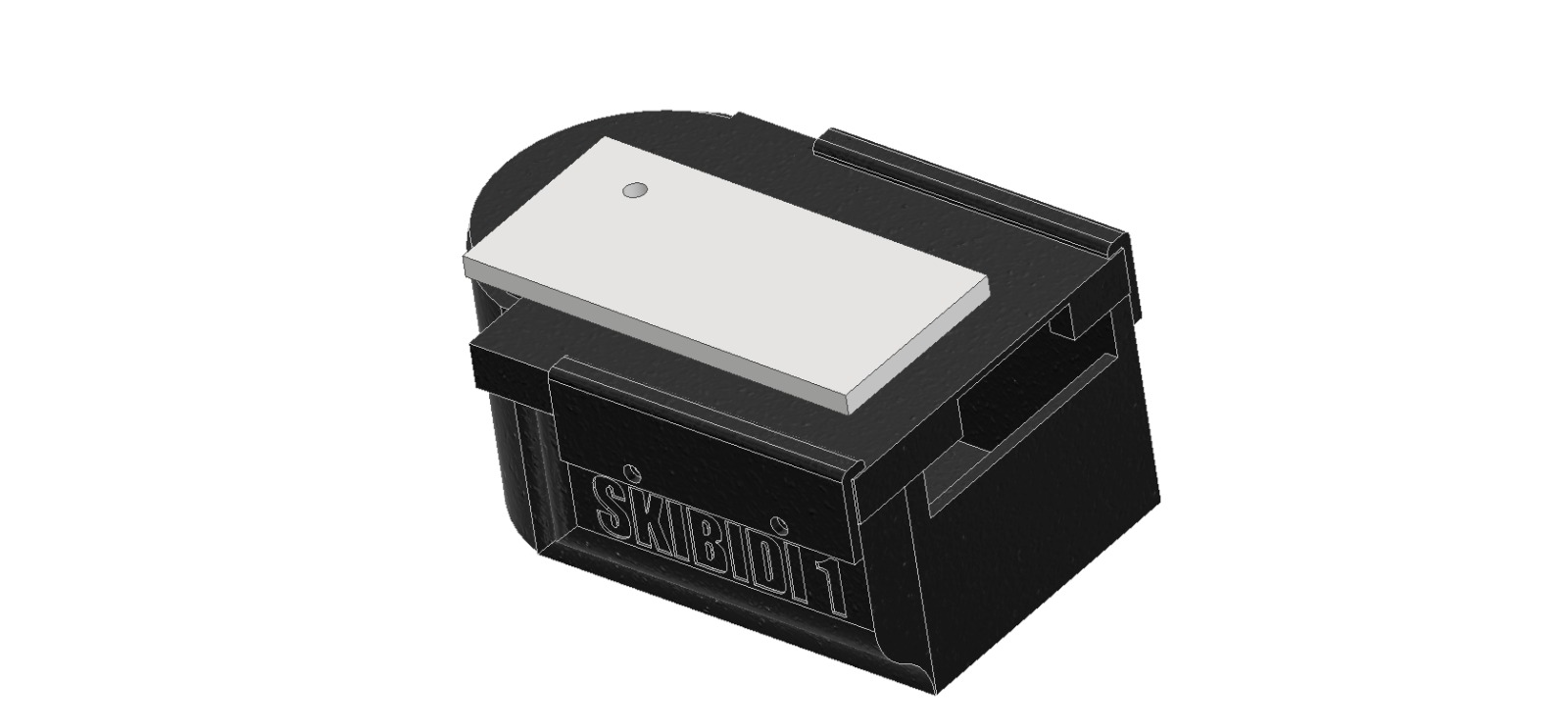}
\caption{Assembled enclosure}
\label{fig:acquisition_assembled}
\end{subfigure}
\hfill
\begin{subfigure}[t]{0.48\textwidth}
\centering
\includegraphics[width=\linewidth]
{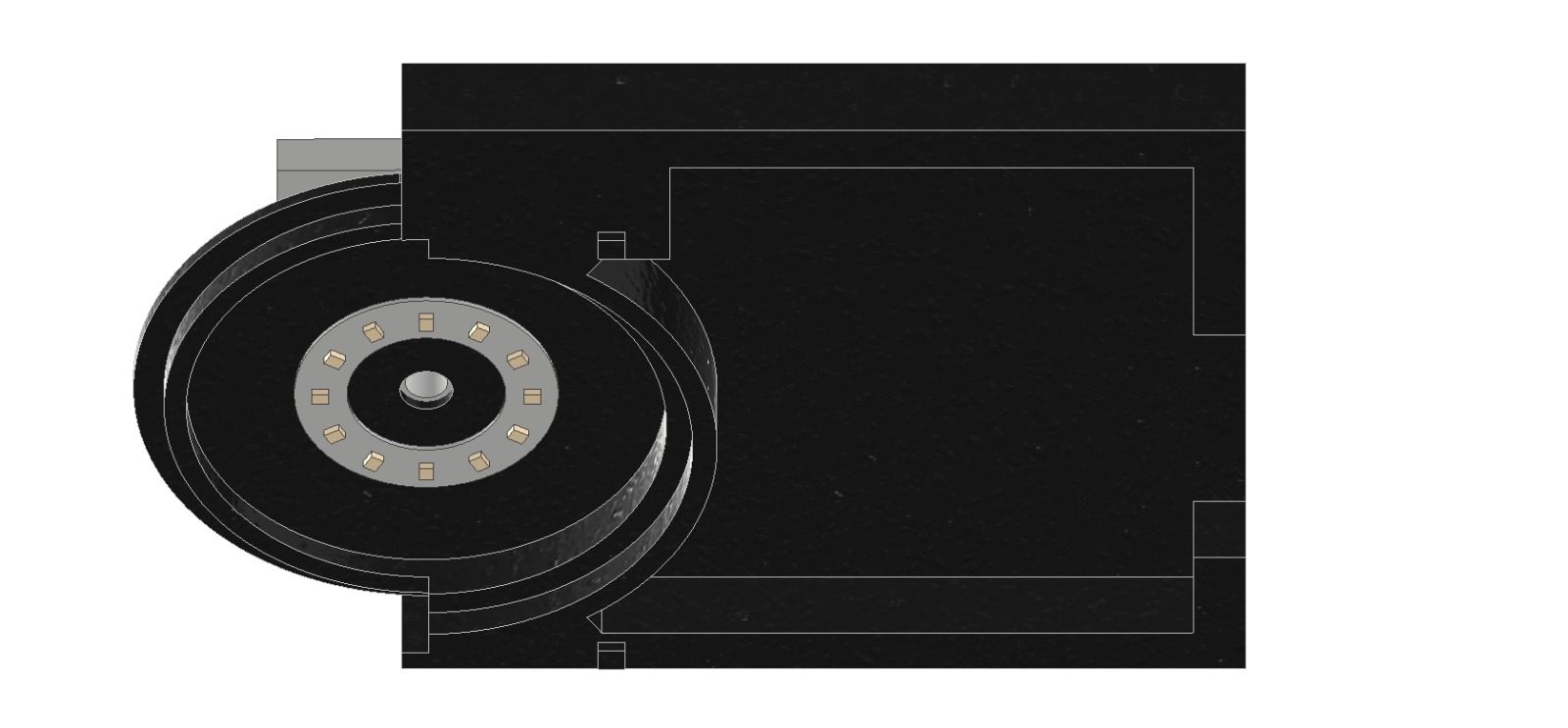}
\caption{Internal LED arrangement}
\label{fig:acquisition_led}
\end{subfigure}
\par\vspace{0.7em}
\begin{subfigure}[t]{0.48\textwidth}
\centering
\includegraphics[width=\linewidth]
{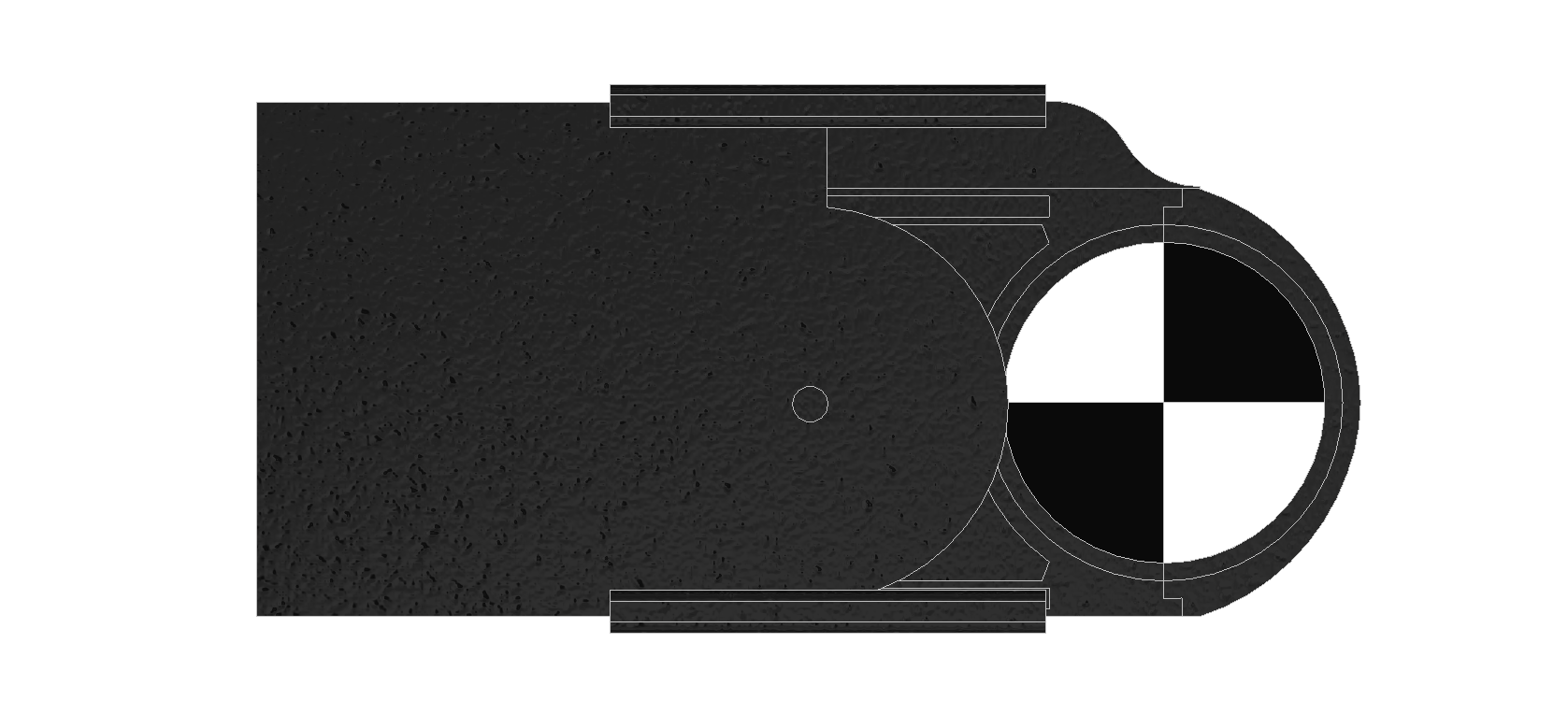}
\caption{Target and imaging geometry}
\label{fig:acquisition_geometry}
\end{subfigure}

\par\vspace{0.7em}

\begin{subfigure}[t]{0.235\textwidth}
\centering
\includegraphics[width=\linewidth]{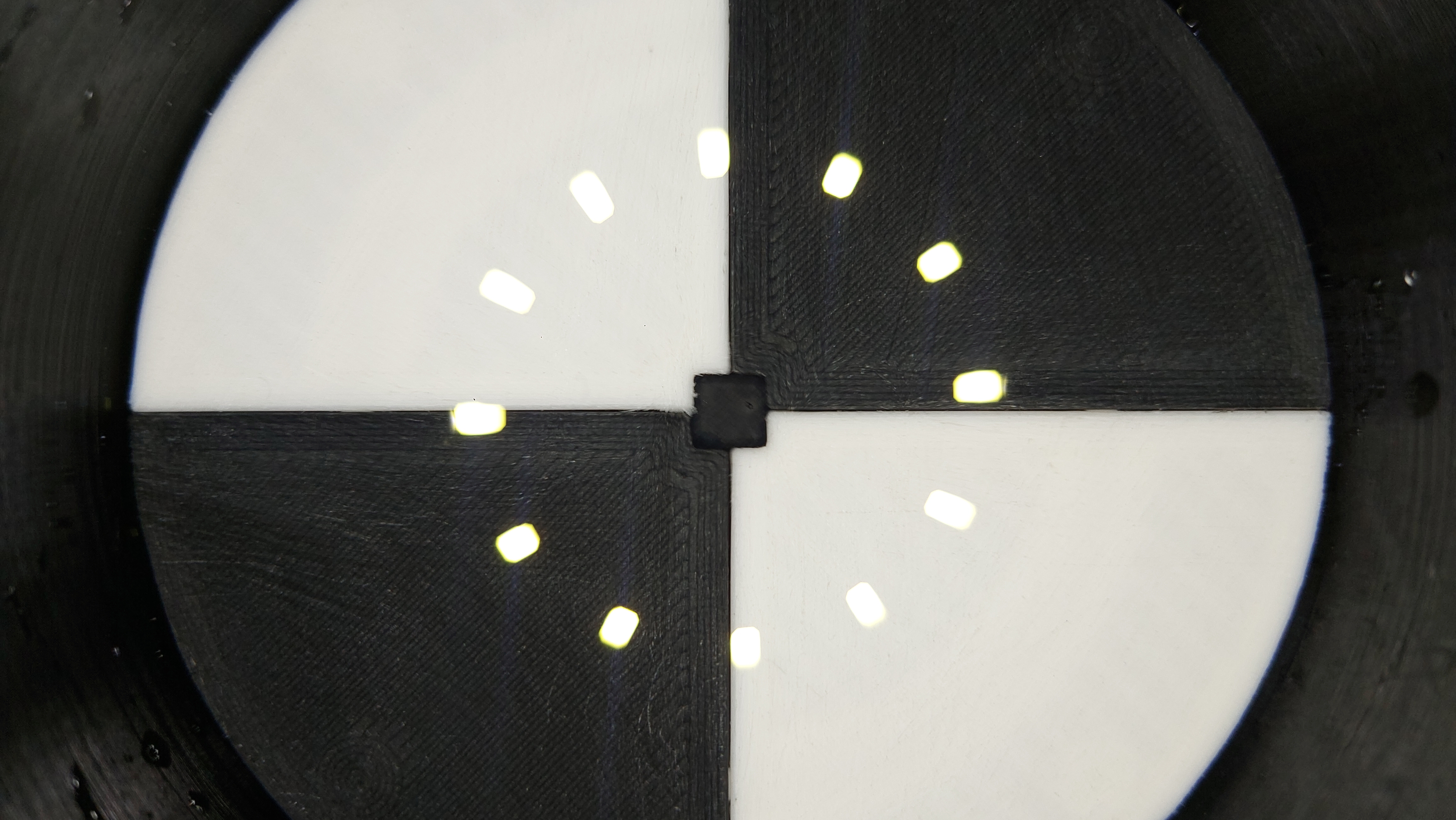}
\caption{0.22 NTU}
\label{fig:acquisition_022}
\end{subfigure}
\hfill
\begin{subfigure}[t]{0.235\textwidth}
\centering
\includegraphics[width=\linewidth]{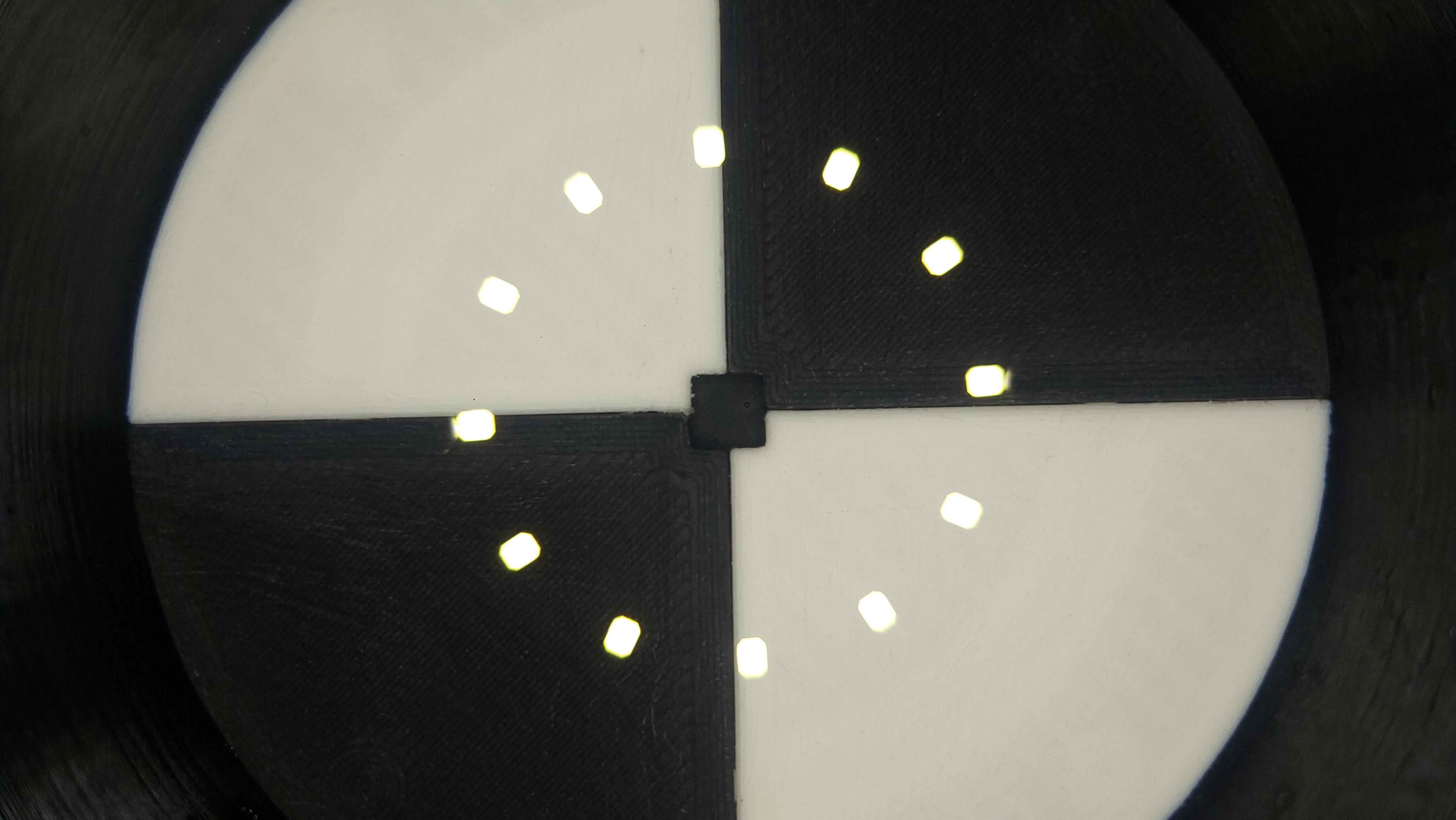}
\caption{5.13 NTU}
\label{fig:acquisition_513}
\end{subfigure}
\hfill
\begin{subfigure}[t]{0.235\textwidth}
\centering
\includegraphics[width=\linewidth]{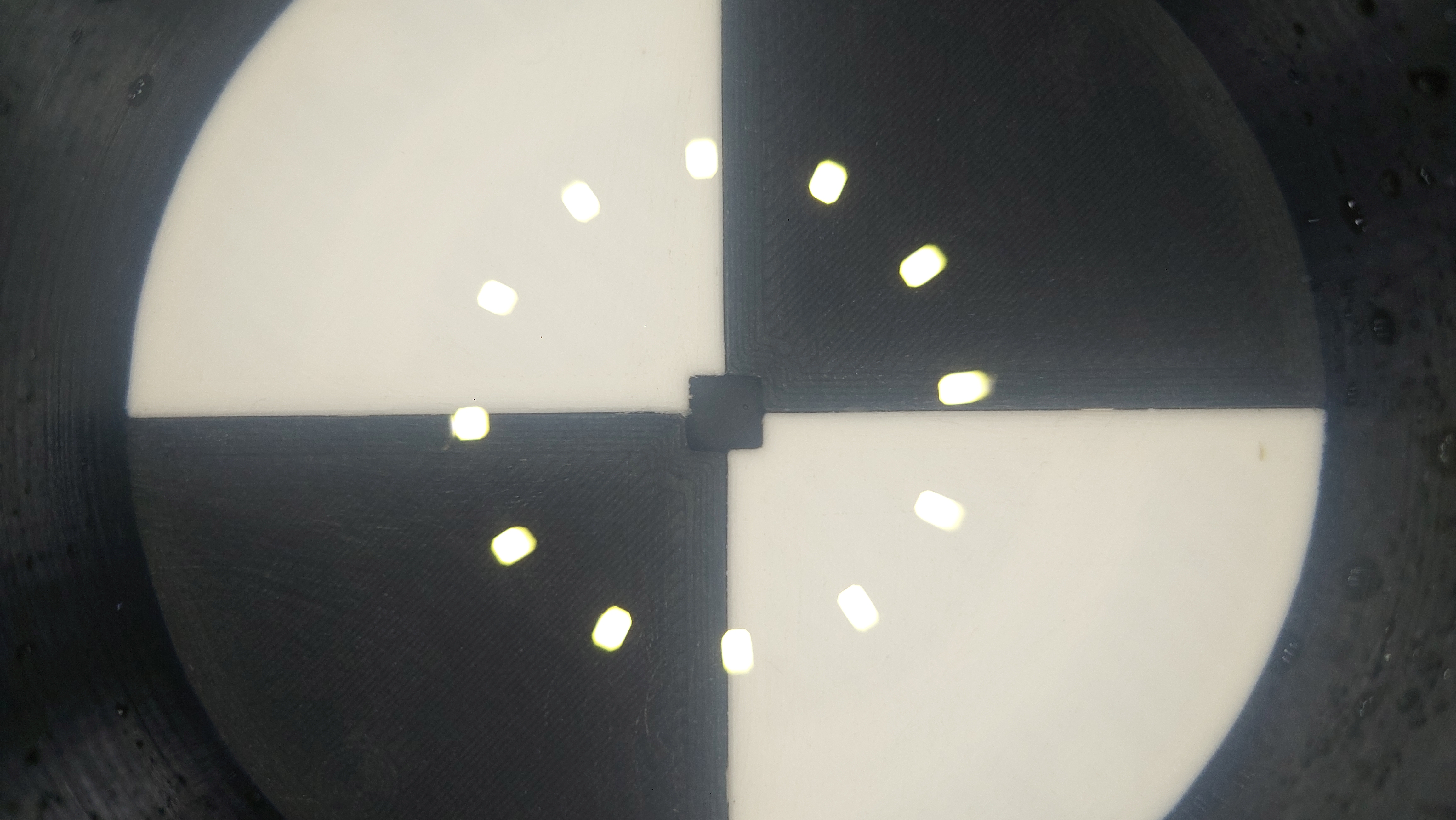}
\caption{17.2 NTU}
\label{fig:acquisition_172}
\end{subfigure}
\hfill
\begin{subfigure}[t]{0.235\textwidth}
\centering
\includegraphics[width=\linewidth]{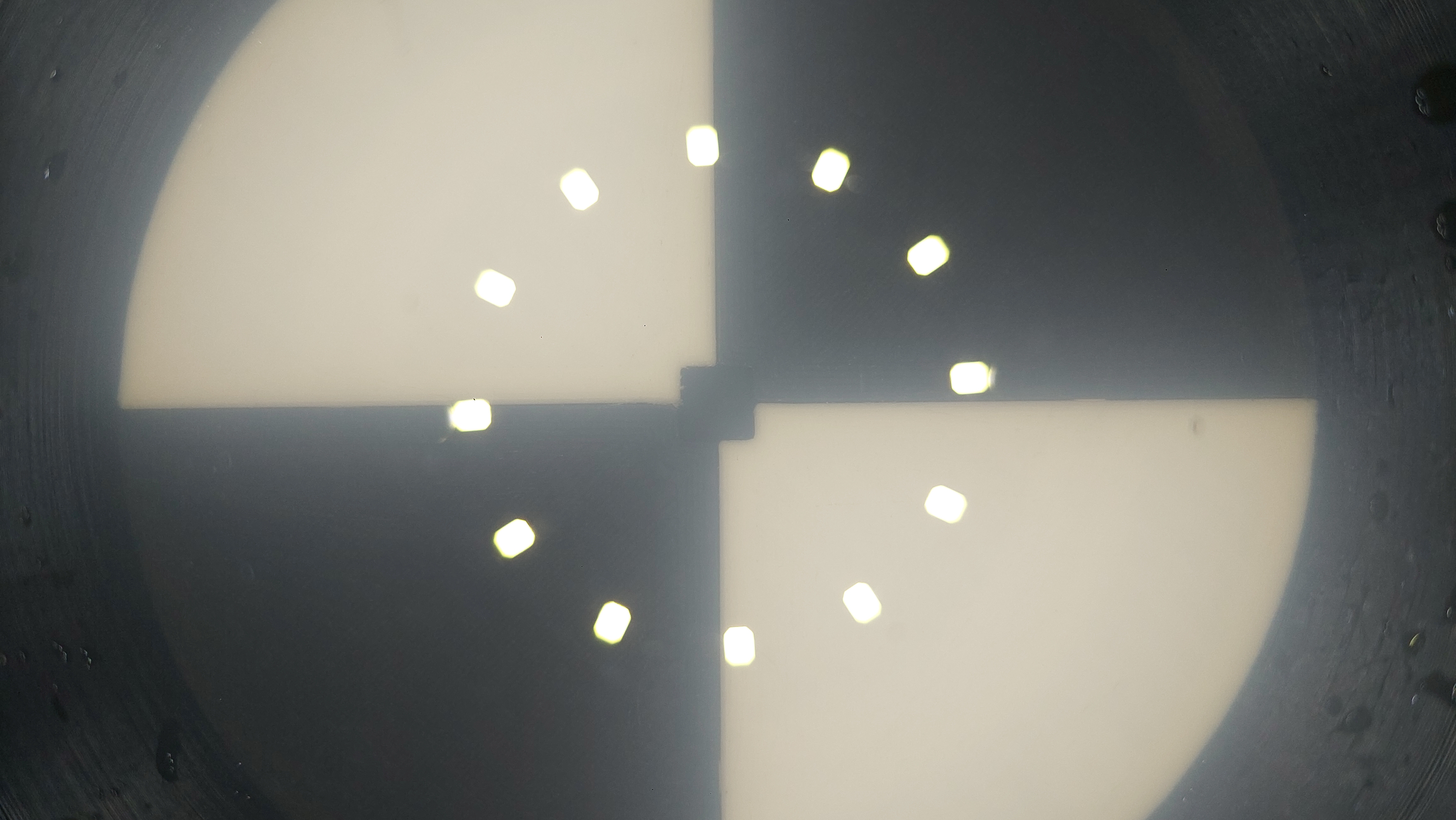}
\caption{49.5 NTU}
\label{fig:acquisition_495}
\end{subfigure}

\caption{\textbf{Controlled-illumination smartphone acquisition system.}
\textbf{A}, CAD representation of the assembled enclosure.
\textbf{B}, Internal arrangement of the twelve-element LED ring surrounding
the imaging region.
\textbf{C}, Top-view geometry showing the circular high-contrast target and
the enclosure structure.
\textbf{D--G}, Representative Samsung acquisitions at reference turbidities
of 0.22, 5.13, 17.2, and 49.5 NTU, respectively. Reference turbidity was
measured before smartphone imaging.}
\label{fig:acquisition}
\end{figure}

Figure~\ref{fig:acquisition} shows the enclosure, illumination arrangement,
optical target, and representative sample images. Table~\ref{tab:hardware}
summarizes the acquisition hardware and recorded camera settings.

\begin{table}[t]
\caption{Acquisition hardware and recorded camera settings. EXIF,
Exchangeable Image File Format.}
\label{tab:hardware}
\centering
\begin{tabular}{@{}ll@{}}
\toprule
Component & Specification \\
\midrule
Primary smartphone & Samsung Galaxy A52 (SM-A525M) \\
Secondary smartphone & Huawei P60 Pro (MNA-LX9) \\
Samsung image resolution & $4624\times2604$ pixels \\
Samsung camera settings & ISO 50, 1/60 s, f/1.8, flash disabled \\
Samsung focal length & 5.23 mm (24 mm full-frame equivalent) \\
Samsung white balance/exposure & Manual white balance and manual exposure \\
Huawei image resolution & $6000\times8000$ pixels \\
Huawei focal length & 6.35 mm (27 mm full-frame equivalent) \\
Huawei exposure settings & ISO 50 or 100; predominantly f/2.5 and 1/100 s \\
Huawei flash and exposure mode & Flash disabled; EXIF-coded automatic exposure \\
Illumination & Fixed internal white LED ring with 12 elements \\
Enclosure & 3D-printed polymer, fixed geometry \\
Container dimensions & Approx. 9 cm diameter and 8.5 cm height \\
Liquid depth & Approx. 4.25 cm \\
Reference instrument & Hanna HI98703-02 turbidimeter \\
\bottomrule
\end{tabular}
\end{table}

All 162 Samsung images had native pixel dimensions of
$4624\times2604$. The recorded acquisition settings were consistent across
the dataset: ISO 50, aperture f/1.8, exposure time 1/60 s, focal length
5.23 mm (24 mm full-frame equivalent), manual white balance, manual exposure,
and flash disabled.

% AUTHOR CHECK: Add verified target dimensions/material, camera-to-target
% distance, LED specifications and supply, and a dimensioned drawing/CAD record.
\subsection{Reference measurement and sample preparation}
\label{subsec:sample_preparation}

Reference turbidity was measured using a Hanna Instruments HI98703-02 portable
ratio-nephelometric turbidimeter, which reports results in NTU. According to
the manufacturer, the instrument method is an adaptation of USEPA Method
180.1 and Standard Methods 2130 B \citep{HannaHI98703Manual}. The instrument
was calibrated at nominal points of 0.10, 15, 100, and 750 NTU. These
calibration points describe the procedure used in the
present study and are not intended as a claim of conformity with ISO
7027-1:2016.

The manufacturer's specified accuracy is $\pm(2\%\ \text{of reading}
+ 0.02\ \text{NTU})$, with a displayed resolution of 0.01 NTU below
10 NTU and 0.1 NTU from 10 to below 100 NTU
\citep{HannaHI98703Manual}. These are instrument specifications, not an
experimentally determined uncertainty budget for the complete smartphone
measurement procedure.
% AUTHOR CHECK: Confirm the measurement mode, calibration/verification
% frequency, standard type and certificate/lot details from laboratory records.
% In particular, verify whether the lowest standard was 0.10 NTU or <0.1 NTU.

Turbidity was generated using powdered milk as a laboratory scattering agent.
Data were acquired in multiple experimental sessions rather than as one
continuous dilution sequence. For each prepared suspension, the
mixture was stirred manually for approximately 20 s. A 10-mL aliquot was then
transferred to the reference turbidimeter vial, and the corresponding
turbidity was recorded in NTU.

Imaging was performed using a cylindrical 3D-printed polymer container with
a diameter of approximately 9 cm and a height of approximately 8.5 cm. The container was filled
to approximately half its height, giving a liquid
depth of approximately 4.25 cm. Because the fill level was set operationally rather than
measured volumetrically for every acquisition, it is reported as approximate.

The reference measurement preceded smartphone imaging. After reference
measurement, the remaining suspension was transferred to the enclosure
container and left undisturbed for approximately 2 min before imaging to
reduce variability caused by residual fluid motion. For samples included in
the paired-device study, both the Samsung and Huawei images were acquired
after the corresponding reference measurement.

\begin{figure}[t]
\centering
\includegraphics[width=\textwidth]{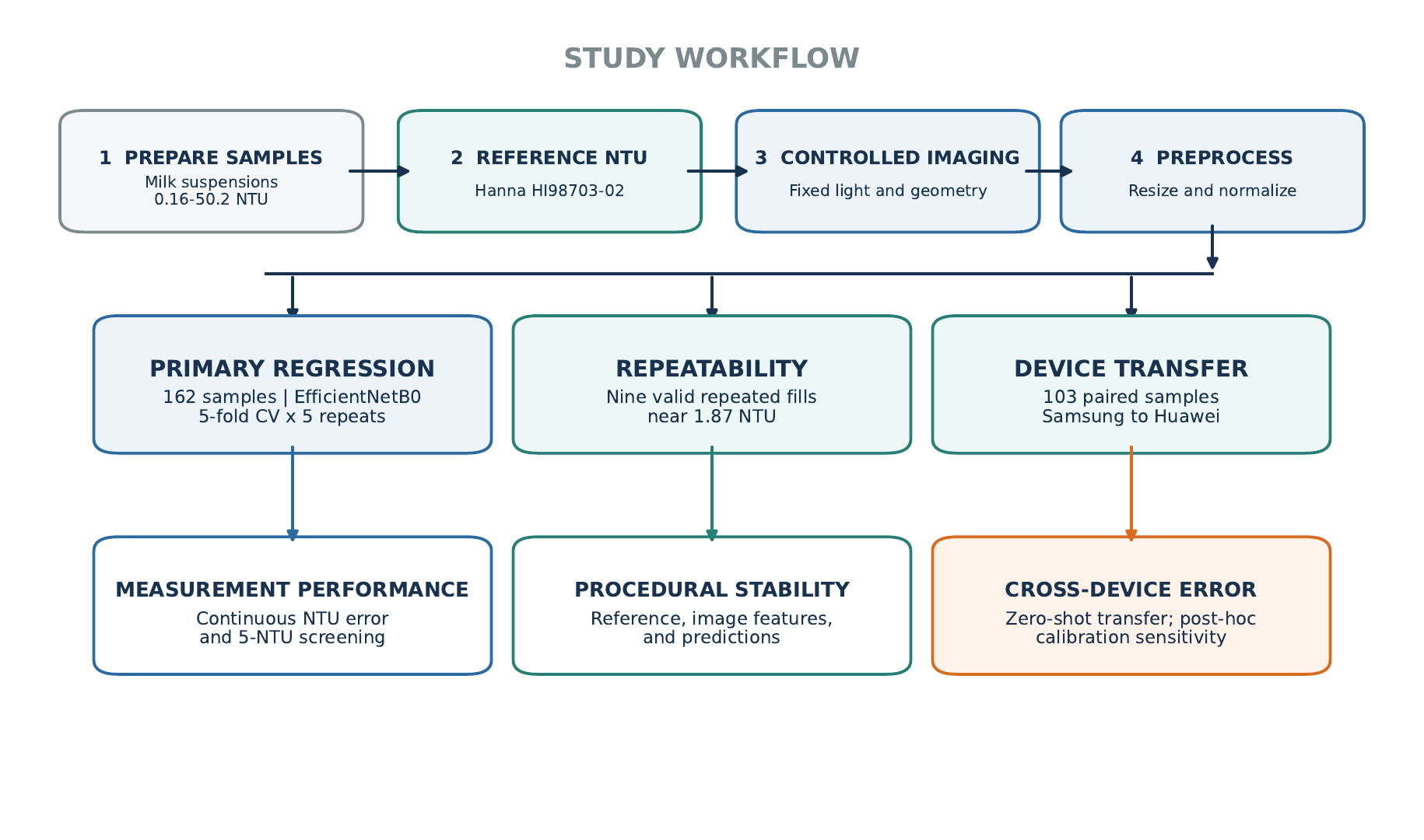}
\caption{Study workflow. Prepared samples were measured with the reference
turbidimeter before controlled smartphone imaging and preprocessing. The
primary analysis evaluated continuous NTU regression using repeated grouped
cross-validation. Separate experiments assessed short-term procedural
repeatability and paired cross-device transfer; device calibration remained a
post-hoc sensitivity analysis. CV, cross-validation.}
\label{fig:workflow}
\end{figure}

Figure~\ref{fig:workflow} summarizes the acquisition and analysis workflow and
distinguishes the primary Samsung regression analysis from the secondary and
post-hoc analyses.

\subsection{Samsung development dataset}
\label{subsec:samsung_dataset}

The Samsung development dataset was acquired in multiple separate
laboratory sessions over approximately three months. It contained 162
physical samples spanning 0.16--50.2 NTU, with 21 samples from 0 to below
1 NTU, 37 from 1 to below 5 NTU, 23 from 5 to below 10 NTU, 30 from 10 to
below 20 NTU, and 51 from 20 to 50.2 NTU. The regression dataset manifest
contained one Samsung image for each physical sample.

Each physical sample was treated as the unit of analysis. The optimization
target was $\log(1+\mathrm{NTU})$. Model outputs were transformed back to
physical units using the inverse transformation and clipped at zero before
calculation of NTU-scale metrics.

\subsection{Image representations and spatial ablation}
\label{subsec:representations}

This secondary experiment was separate from the primary continuous-regression
analysis. It used the reference turbidimeter value to assign each Samsung
sample to a binary class below 5 NTU or at or above 5 NTU. The 5-NTU value was
therefore an analysis threshold applied after reference measurement, not part
of the measurement procedure. The experiment used \emph{spatial ablation}:
selected image regions were retained or removed while the learning and
evaluation procedure was held fixed. This tested how much classification
information was supplied by the submerged target, the illuminated regions,
and the wider enclosure context.

The full-image representation used the complete acquisition image after
resizing. The central-target representation first extracted a centered square
crop with a side length equal to 94\% of the shorter original image dimension.
The light-neighborhood representation used the same target crop and retained
twelve fixed circular neighborhoods centered on the visible illumination
elements. The neighborhood centers were specified in normalized,
model-oriented crop coordinates. Each neighborhood had a radius of 0.065 of
the normalized crop dimension and a feathered boundary width of 0.008.
Pixels outside the selected neighborhoods were replaced by a neutral value of
127.5 before ImageNet-compatible preprocessing. The twelve normalized center
coordinates and complete preprocessing configuration will be provided with
the released analysis code.

\begin{figure}[t]
\centering
% ARTWORK REVIEW: Check panel-label readability in the final compiled PDF.
\includegraphics[width=\textwidth]{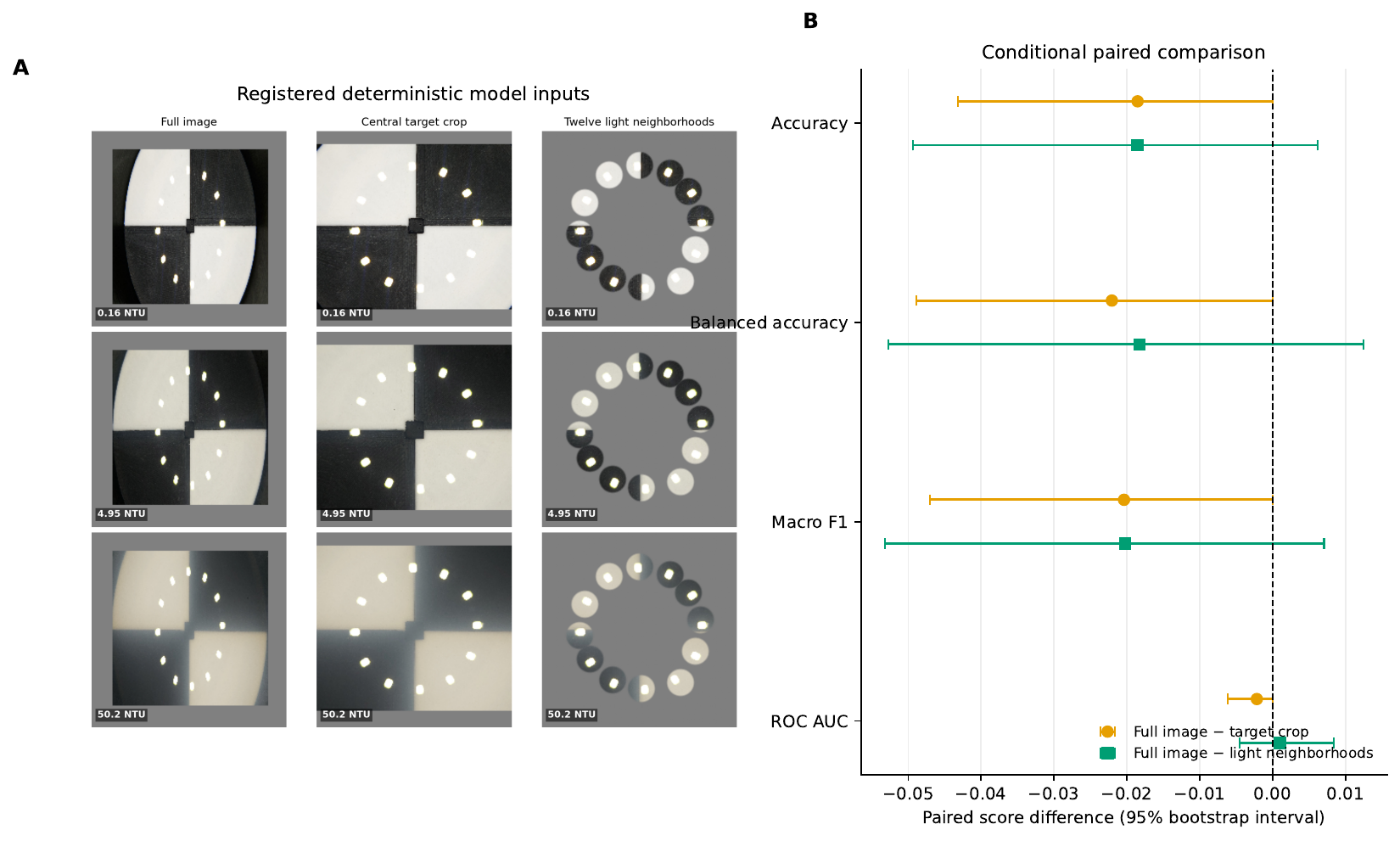}
\caption{Registered spatial representations and paired Samsung classification
performance. \textbf{A}, Representative Samsung samples at 0.16, 4.95, and
50.2 NTU shown after deterministic geometric registration as the full image,
central target crop, and twelve fixed light neighborhoods. \textbf{B}, Paired
differences in sample-level classification performance, calculated as full
image minus each restricted representation. Points show the observed
differences and horizontal bars show 95\% paired, stratified physical-sample
bootstrap intervals conditional on the fitted repeated cross-validation
models. Full-image-minus-target-crop estimates were negative, with intervals
reaching zero at their upper boundary, whereas comparisons between the full
image and light neighborhoods were inconclusive. The target crop therefore had
the highest observed point estimates, but the analysis supports at most a
modest in-domain classification advantage and does not establish superiority
for continuous regression or device transfer.}
\label{fig:ablation}
\end{figure}

The three representations are illustrated in
Figure~\ref{fig:ablation}A: the full acquisition image, a centered square
target crop, and twelve fixed light neighborhoods. All configurations used matched physical
samples and the same repeated grouped cross-validation protocol. Detailed
classification metrics and paired comparisons are reported in the
Supplementary Information.

Visual inspection of the representation panels identified variation in the
apparent target center and scale among acquisition images. A deterministic
registration sensitivity analysis was therefore performed after the primary
analysis. Illumination-element geometry was detected without using turbidity
labels, and each Samsung image was translated and isotropically rescaled to a
common canonical geometry. Detection succeeded for all 162 images. The full
image, central target crop, and twelve-light-neighborhood representations were
then reevaluated using the registered inputs. In addition to the binary
classification ablation, continuous-NTU regression was repeated for all three
registered representations using the same samples, folds, random seeds for each repetition,
model architecture, and training settings. Because registration was introduced
after visual inspection of the figure panels, these results are treated as a
secondary sensitivity analysis rather than a replacement for the prespecified
full-image regression.

\subsection{Regression model and training}
\label{subsec:model_training}

The primary continuous-turbidity model used EfficientNetB0 initialized with
ImageNet-pretrained weights. JPEG images were decoded as three-channel images,
resized to $224\times224$ pixels using bilinear interpolation, and processed
using the corresponding Keras EfficientNet preprocessing function. The full
acquisition image was used without a spatial crop for the primary regression
analysis. Data augmentation was disabled.
This choice preserved the physically fixed apparatus geometry and avoided
introducing synthetic rotations, color changes, or displacements that were not
part of the recorded acquisition protocol.

The EfficientNetB0 feature extractor was followed by global average pooling,
dropout with probability 0.2, a dense layer with 256 rectified linear units,
and a scalar linear regression output. Optimization was performed on
$\log(1+\mathrm{NTU})$ using the Adam optimizer and mean squared error loss.
Predictions were transformed back to physical units according to
$\widehat{\mathrm{NTU}}=\exp(\hat{y})-1$, and negative values following the
inverse transformation were clipped at zero.

Training used a batch size of 16 and an initial learning rate of $10^{-3}$.
The pretrained backbone was initially frozen while the regression head was
optimized for a maximum of 60 epochs. Early stopping monitored validation
loss with a patience of eight epochs and restored the weights corresponding to
the minimum validation loss. The final 20 layers of the EfficientNetB0
backbone were subsequently made trainable, except for batch-normalization
layers, which remained frozen. Fine-tuning used Adam with a learning rate of
$10^{-5}$ for a maximum of 20 additional epochs and the same early-stopping
criterion. The base random seed was 42.

\subsection{Repeated grouped cross-validation}
\label{subsec:grouped_cv}

The primary Samsung evaluation used five-fold grouped cross-validation
repeated five times. In each repetition, the physical samples were divided
into five non-overlapping subsets. Four subsets were used for model development
and the remaining subset was held out for evaluation; this process was cycled
until every sample had been evaluated once. The entire five-fold partition was
then regenerated four more times, giving each sample five predictions from
models that had not been trained on that sample. The physical-sample identifier
was the grouping variable, ensuring that all captures associated with a sample
would remain in the same subset. Although the regression dataset manifest
contained one image per physical sample, grouping was retained to prevent images of the same sample from appearing
in both the development and evaluation subsets if repeated captures were
added in future studies.

For each repetition, samples were assigned to outer folds using
StratifiedGroupKFold with shuffling. Stratification used five quantile-based
bins of reference NTU. Within each outer development set, 15\% of the physical
samples were reserved as an inner tuning subset using a stratified random
split based on up to five NTU quantile bins. Neural-network fitting and early
stopping used only the remaining training samples and the inner tuning subset.
The outer held-out fold was used only for evaluation.

The raw JPEG images remained the model inputs; the folds only determined which
physical samples were used for development or held-out evaluation. The outer
partitions used reproducible seeds derived from the base seed of 42. Each
physical sample received one held-out prediction in every repetition, and its
primary sample-level estimate was the median of the five repeated out-of-fold
predictions. Fold-level results are correlated because their development sets
overlap; fold dispersion was therefore treated as descriptive rather than as
independent experimental replication.

Grouping by physical sample does not establish independence across
acquisition sessions or stock preparations. This evaluation was not designed
as a leave-session-out or leave-batch-out study and therefore does not
establish performance on a new session or independently prepared batch.
% AUTHOR CHECK: Report session counts and whether samples shared a stock
% preparation; confirm how sessions and preparation batches span outer folds.

\subsection{Low-NTU and operational-threshold analysis}\label{subsec:low_ntu_methods}

The predefined ranges were 0--$<$1, 1--$<$5, 5--$<$10, 10--$<$20, and
20--50.2 NTU. The operational threshold was 5 NTU, with categories labeled
``below 5 NTU'' and ``at or above 5 NTU.''

Numerical error was summarized separately within each predefined NTU range.
Continuous predictions were also converted to binary categories at 5 NTU. Accuracy was defined
as the proportion of correctly classified samples; sensitivity was the
proportion of reference samples at or above 5 NTU predicted on that side of
the threshold; and specificity was the proportion of reference samples below
5 NTU predicted below the threshold. Balanced accuracy and macro-F1 were also
reported. A false-negative error was defined as a reference value at or
above 5 NTU with a prediction below 5 NTU, whereas a false-positive error
was defined as a reference value below 5 NTU with a prediction at or above
5 NTU. The seven samples between 4 and 6 NTU formed a descriptive boundary
subset and were not treated as an independent validation sample.

\subsection{Repeatability experiment}
\label{subsec:repeatability_methods}

The repeatability experiment was performed separately from the three-month
development dataset as one continuous acquisition session. Ten successive
fills were obtained from one bulk suspension with reference turbidity near
1.87 NTU. The experiment therefore evaluated repeatability of the combined
refill, reference-measurement, settling, imaging, feature-extraction, and
prediction procedure under one low-turbidity condition; it was not designed
to estimate total measurement uncertainty across the full NTU range.

Nine fills constituted the prespecified primary analysis. The first
observation was retained in the audit trail but excluded from the primary
summary because of an independently confirmed acquisition or sample-labeling
provenance error. Exclusion was determined without reference to model output.
A sensitivity analysis including all ten observations was performed separately.

Optical-feature variability was summarized across the nine valid fills. The
three neural-network outputs evaluated in this experiment were probabilities
from the earlier binary-classification models: a small convolutional neural
network, EfficientNetB0, and ConvNeXt-Tiny. These were classification
probabilities rather than continuous NTU estimates. Prediction consistency
was therefore evaluated with respect to the below-5-NTU category. This
single-level experiment cannot be extrapolated to other turbidity values or
used as a full measurement-uncertainty budget. Because continuous-NTU
regression predictions were not evaluated in this experiment, it does not
quantify repeatability of the regression output in NTU.

\subsection{Paired Samsung--Huawei device-transfer evaluation}
\label{subsec:device_transfer_methods}

The paired-device dataset contained Samsung Galaxy A52 and Huawei P60 Pro
(MNA-LX9) images of the same 103 physical samples, spanning 0.16--50.2 NTU.
All entries in the pairing manifest were explicitly identified as the same
physical sample photographed by both phones. The pairing table linked each
physical-sample identifier and reference NTU value to one Samsung filename
and one Huawei filename. The reference turbidimeter measurement preceded
both smartphone images.

%The Huawei images were acquired in four sessions between 7 and 31 July 2025.
The Huawei images were acquired across four experimental sessions.
All 103 images had a native resolution of $6000\times8000$ pixels, a focal
length of 6.35 mm (27 mm full-frame equivalent), and were captured with the
flash disabled. ISO was 100 for 60 images and 50 for 43 images. For 102 of
the 103 images, the recorded aperture was f/2.5, exposure time was 1/100 s,
and white balance was EXIF-coded as manual. One image used f/2.0 with
EXIF-coded automatic white balance, and a different image used an exposure
time of 1/125 s. All images were retained because these variations were
identified during the retrospective metadata audit and were not associated
with an independently documented acquisition failure.

The Huawei EXIF metadata contained two software builds:
\texttt{MNA-L29 14.2.0.175(C605E2R3P1)} for 52 images and
\texttt{MNA-L29 14.2.0.176(C605E2R3P1)} for 51 images. The handset itself was
identified as model MNA-LX9, whereas the EXIF \texttt{Software} field reported
MNA-L29 build identifiers. Consequently, the paired-device analysis evaluates
transfer across the complete observed Huawei acquisition pipeline, including
its camera sensor, optics, software processing, and recorded
acquisition-setting variation. 

For each outer cross-validation fold, an EfficientNetB0 regression model was
trained and tuned exclusively using Samsung images from the outer development
set. The fitted fold model was then applied, without Huawei-specific
neural-network fitting or fine-tuning, to the paired Samsung and Huawei images
belonging to held-out physical samples. Samsung and Huawei images were
processed using the same deterministic resizing, normalization, and spatial
representation. Huawei images were not used for training, early stopping,
preprocessing selection, or original architecture selection.

Evaluation reused the exact held-out assignments from the repeated grouped
Samsung analysis described above. Each paired sample therefore received five
predictions per device from models for which that physical sample had been
held out, and these predictions were aggregated using the median.

Both the full-image and centered target-crop representations were evaluated.
The target crop used a centered square with a side length equal to 94\% of
the shorter original image dimension. The uncalibrated Huawei predictions
constituted the primary device-transfer result. Because the preference for
the full-image representation was informed by inspection of the paired-device
results, it is reported as an observed representation-dependent transfer
finding rather than as a prospectively selected configuration.

The neural-network image loader decoded the stored JPEG pixel arrays without
applying EXIF orientation metadata. An acquisition audit subsequently showed
that all Samsung images carried EXIF orientation value 6, whereas the Huawei
images carried the non-standard value 0. Apparatus landmarks indicated that
rotating the Huawei stored pixels by $90^\circ$ counterclockwise aligned them
with the stored Samsung pixel orientation used during model training. We
therefore performed a post-hoc orientation sensitivity analysis in which the
Huawei images were rotated before the otherwise unchanged crop, resize, and
normalization operations. Raw and orientation-matched Huawei images were
evaluated using the same fitted Samsung models. This analysis was not used for
neural-network training or original model selection.

\subsection{Post-hoc cross-fitted Huawei calibration}
\label{subsec:calibration_methods}

Three post-hoc calibration approaches were evaluated. The non-negative
multiplier used $\widehat{y}_{\mathrm{cal}}=a\widehat{y}$ with $a\geq0$ and
no intercept, so it could correct only an overall scale difference. The
log-affine method fitted an intercept and slope after transforming both the
prediction and reference value with $\log(1+y)$, allowing both offset and
scale correction while reducing the influence of the largest NTU values.
Isotonic regression fitted a flexible non-decreasing mapping without assuming
a particular algebraic form; it can capture nonlinear calibration behavior
but has greater overfitting risk in a small dataset.

Calibration was cross-fitted using five folds and five repetitions. For each
held-out physical sample, the calibrator was fitted using only the reference
NTU values and Huawei predictions of the remaining calibration samples. A
sample's own reference NTU was therefore never used to fit the calibrator
applied to that sample.

Within this explicitly post-hoc analysis, the log-affine mapping was
designated as the primary comparison method. The multiplier served as a
simple benchmark, while isotonic regression was treated as an exploratory
flexible method. Although cross-fitting prevents direct sample leakage, the
calibration analysis and candidate mappings were developed after inspection
of the zero-shot transfer behavior. Consequently, this analysis does not
constitute an independent prospective device-calibration study.

\subsection{Measurement model and uncertainty considerations}
\label{subsec:measurement_uncertainty}

The measurand modeled in this study was the turbidity value reported in NTU by
the calibrated reference turbidimeter for each prepared suspension. The
smartphone result was an estimate of that reference value obtained from the
stored image after the specified preprocessing and regression procedure. The
reported prediction errors therefore characterize agreement with the reference
instrument under the tested preparation and acquisition conditions; they are
not a determination of an independently realized true turbidity value.

Several components can contribute to the observed disagreement: reference-instrument
repeatability and calibration, preparation and aliquoting of the
suspension, the interval between reference measurement and imaging, settling
and refill variation, illumination and geometry, camera acquisition and image
processing, and fitted-model variability. The repeated grouped cross-validation
analysis captures sample-level predictive variation within the Samsung dataset,
and the paired-device experiment quantifies the combined effect of changing the
camera and its acquisition pipeline for the same physical samples. The
single-level repeatability experiment provides a local estimate of short-term
procedural variation near 1.87 NTU. Because repeated reference and image
measurements were not acquired across the full range, these components cannot
be separated into a complete measurement-uncertainty budget. Accordingly,
intervals reported in this work describe statistical uncertainty conditional
on the observed samples and fitted analysis procedure, rather than expanded
measurement uncertainty or metrological traceability across the full range.

\subsection{Statistical analysis}\label{subsec:statistics}

Primary continuous-regression metrics were mean absolute error (MAE), root
mean squared error (RMSE), median absolute error, mean signed error
(bias, defined as prediction minus reference NTU), coefficient of
determination ($R^2$), Spearman's rank correlation coefficient, and MAE on the
$\log(1+\mathrm{NTU})$ scale. Metrics were calculated using the median repeated
out-of-fold prediction for each physical sample.

MAE describes the average absolute NTU error, whereas RMSE gives greater
weight to large errors. Bias indicates systematic overprediction or
underprediction. $R^2$ describes how much of the variation in reference NTU is
captured by the predictions, and Spearman's $\rho$ measures whether samples
are ranked in the correct turbidity order even when the relationship is not
perfectly linear. For thresholded predictions, balanced accuracy gives equal
weight to the two reference classes and macro-F1 averages class-specific F1
scores.

Uncertainty intervals for continuous metrics were obtained from 5000
physical-sample bootstrap repetitions using random seed 42. In each repetition, the observed
physical samples were drawn with replacement and the metric was recalculated;
the resulting empirical distribution was used to obtain the interval.
Resampling was stratified by the predefined reference ranges 0--$<$1, 1--$<$5,
5--$<$10, 10--$<$20, and 20--50.2 NTU. Comparisons between spatial
representations and paired devices used paired resampling so that matched
predictions from the same physical sample remained together. These intervals
quantify physical-sample uncertainty conditional on the existing fitted
cross-validation models and fixed analysis procedures; they do not include
neural-network training variability.

At the 5-NTU operational threshold, uncertainty intervals for accuracy,
balanced accuracy, and macro-F1 were obtained using the stratified
physical-sample bootstrap described above. Sensitivity for reference samples
at or above 5 NTU and specificity for reference samples below 5 NTU were
reported with two-sided 95\% Wilson score intervals. The Wilson method is a
confidence interval for a binomial proportion and is more stable than the
simple normal approximation when a class contains relatively few samples or
the observed proportion is close to zero or one.

% ============================================================================
\section{Results}\label{sec:results}

\subsection{Primary Samsung continuous-NTU regression}\label{subsec:samsung_regression_results}

\begin{figure}[t]
\centering
\includegraphics[width=\textwidth]{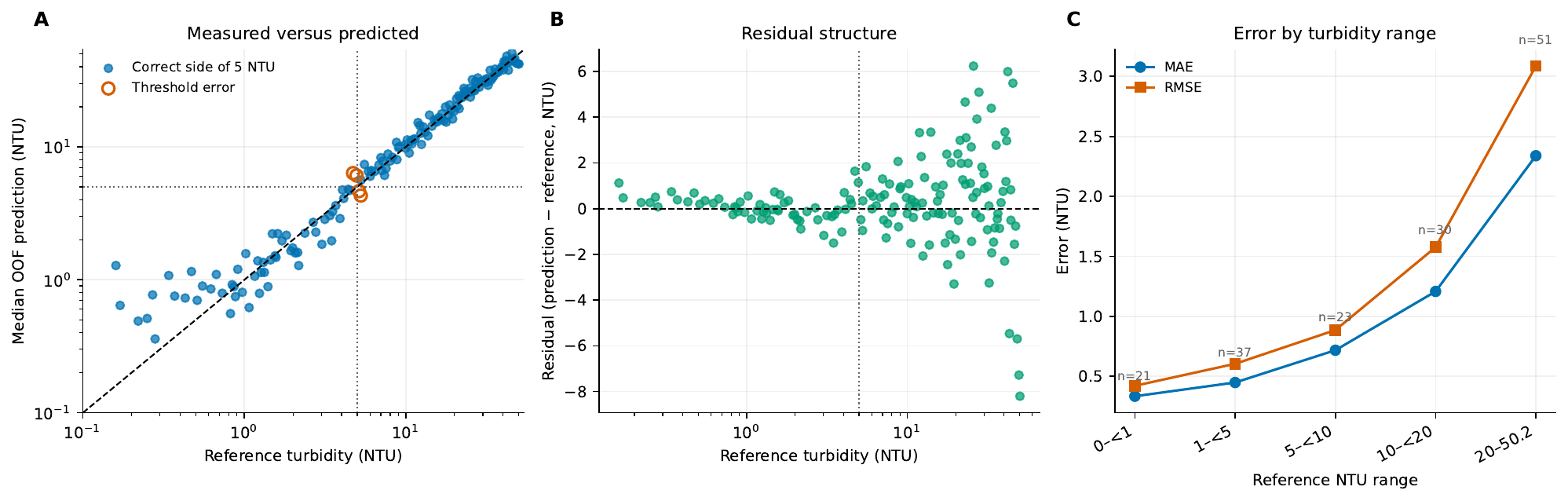}
\caption{Repeated grouped cross-validated Samsung continuous-NTU regression.
\textbf{A}, Reference turbidity versus the median of five repeated out-of-fold
predictions for each of 162 physical samples. The dashed line indicates
identity, the dotted lines indicate the 5-NTU operational threshold, and open
orange markers identify threshold-side errors. \textbf{B}, Residuals, defined
as predicted minus reference NTU, versus reference turbidity. \textbf{C}, MAE
and RMSE point estimates in the five predefined turbidity ranges; sample counts
are shown above the corresponding estimates.}
\label{fig:samsung_regression}
\end{figure}

Across 162 physical samples, the median of the five repeated out-of-fold
EfficientNetB0 predictions for each sample yielded
an MAE of 1.208 NTU (95\% interval 1.018--1.407), RMSE of 1.918 NTU
(1.572--2.246), median absolute error of 0.635 NTU (0.495--0.811), and
$R^2=0.9816$ (0.9757--0.9872). Estimated bias was 0.243 NTU
($-0.052$--0.524), and Spearman's $\rho$ was 0.9921 (0.9878--0.9943).
In practical terms, the model's predictions differed from the reference by
about 1.2 NTU on average across the full 0.16--50.2 NTU range, captured more
than 98\% of the observed variance, and preserved the ordering of samples very
closely. These values indicate strong internal predictive performance under
the tested acquisition conditions, but do not constitute independent
external validation. Figure~\ref{fig:samsung_regression} shows the predictions
and residuals; Table~\ref{tab:regression_ranges} reports error by turbidity
range.

\begin{table}[t]
\caption{Samsung regression error by predefined NTU range.}
\label{tab:regression_ranges}
\centering
\begin{tabular}{@{}lrrrr@{}}
\toprule
Reference range & $n$ & MAE & RMSE & Bias \\
\midrule
0--$<$1 NTU & 21 & 0.335 & 0.421 & 0.281 \\
1--$<$5 NTU & 37 & 0.449 & 0.605 & $-0.079$ \\
5--$<$10 NTU & 23 & 0.718 & 0.886 & 0.277 \\
10--$<$20 NTU & 30 & 1.208 & 1.577 & 0.111 \\
20--50.2 NTU & 51 & 2.339 & 3.086 & 0.523 \\
\bottomrule
\end{tabular}
\end{table}

\subsection{Spatial-representation comparison}
\label{subsec:ablation_results}

After deterministic geometric registration, the central target crop had the
highest observed Samsung binary-classification point estimates, followed
closely by the twelve light neighborhoods; the full image had lower observed
point estimates (Figure~\ref{fig:ablation} and Supplementary Table S2). For the
target-crop comparison, the paired intervals for full image minus target crop
extended to zero at their upper boundary. Differences between the target crop
and light neighborhoods were inconclusive. These findings indicate at most a
modest registered in-domain classification advantage for the target crop and
do not establish a single universally preferable representation.

The corresponding registered continuous-regression sensitivity analysis gave
a different ordering (Supplementary Table S3). The full image had the lowest
observed MAE (1.250 NTU), lowest RMSE (1.791 NTU), and highest $R^2$ (0.9840).
The target crop yielded an MAE of 1.353 NTU, RMSE of 2.024 NTU, and
$R^2=0.9795$. Paired target-crop-minus-full differences were 0.103 NTU for MAE
(95\% interval $-0.102$--0.310), 0.233 NTU for RMSE ($-0.118$--0.571), and
$-0.0045$ for $R^2$ ($-0.0113$--0.0024); thus, the registered regression
analysis did not establish a clear difference between these two
representations. The twelve-light representation had higher error than the
full image: paired differences were 0.416 NTU for MAE (0.171--0.666) and
0.540 NTU for RMSE (0.136--0.950), while the $R^2$ difference was $-0.0111$
($-0.0198$--$-0.0027$).

Range-specific point estimates showed that the target crop had slightly lower
MAE than the full image below 5 NTU (0.337 versus 0.420 NTU below 1 NTU and
0.591 versus 0.608 NTU from 1 to below 5 NTU), whereas the full image had lower
MAE in each predefined range at or above 5 NTU. These range-specific values are
descriptive. The sensitivity analysis therefore supports retaining the
full-image model as the primary continuous-regression analysis while showing
that representation ranking depends on the prediction task and turbidity
range.

The preliminary Samsung architecture comparison is reported separately in
Supplementary Table S1 because it used an earlier binary-classification task
and a single held-out test split rather than the repeated continuous-regression
evaluation.

\subsection{Low-NTU accuracy and the 5-NTU operating threshold}\label{subsec:low_ntu_results}

For the primary Samsung full-image regression, MAE was 0.335 NTU below 1 NTU,
0.449 NTU from 1 to below 5 NTU, and 0.718 NTU from 5 to below 10 NTU.
Across all 162 samples, the 5-NTU threshold accuracy was 0.9753
(95\% interval 0.9506--0.9938), balanced accuracy was 0.9731
(0.9435--0.9952), and macro-F1 was 0.9731 (0.9459--0.9933).

\begin{figure}[t]
\centering
% ASSET CORRECTION REQUIRED: Update artwork labels to False negative at 5 NTU and False positive at 5 NTU; remove unsafe miss and conservative flag.
\includegraphics[width=\textwidth]{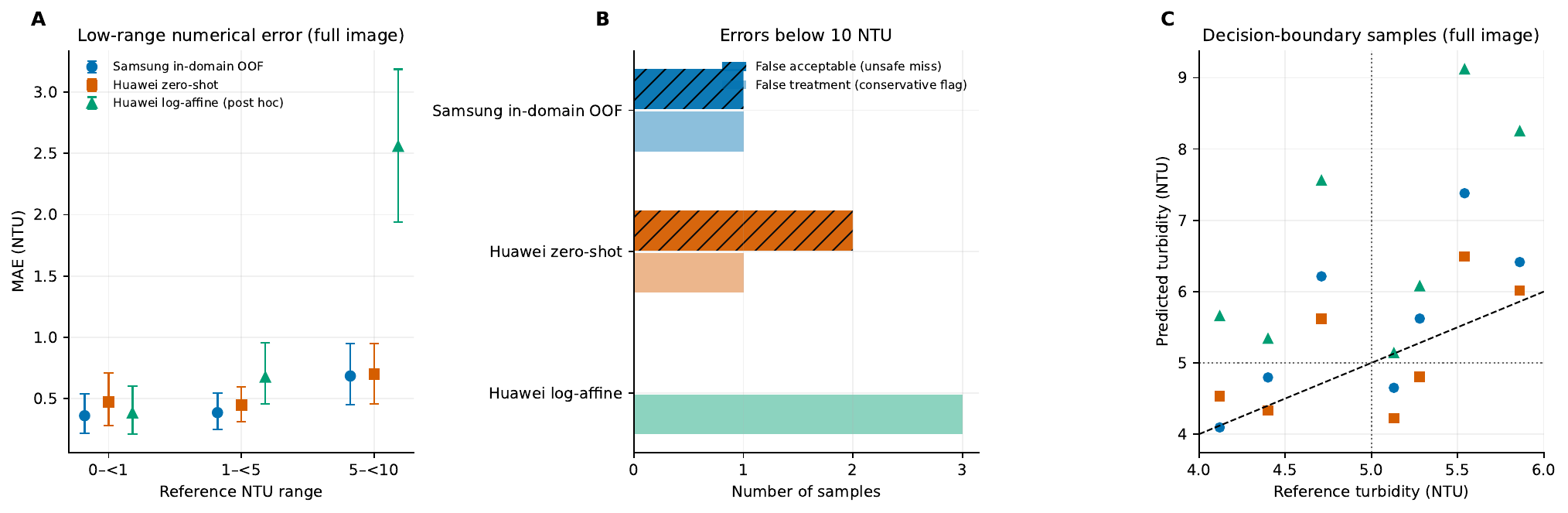}
\caption{Full-image low-NTU performance in the 103-sample paired
Samsung--Huawei device-transfer subset. \textbf{A}, MAE and 95\% stratified physical-sample bootstrap
intervals in the 0--$<$1, 1--$<$5, and 5--$<$10 NTU ranges for Samsung
in-domain out-of-fold prediction, Huawei zero-shot prediction, and cross-fitted
Huawei log-affine calibration. \textbf{B}, Numbers of false-negative errors
(reference NTU at or above 5 but prediction below 5) and false-positive errors
(reference NTU below 5 but prediction at or above 5) among the 52 samples below
10 NTU. \textbf{C}, Predictions for the seven physical samples in the
descriptive 4--6 NTU boundary subset. Dashed and dotted lines indicate identity
and the 5-NTU threshold, respectively. Calibration was defined post hoc and is
shown as a secondary exploratory analysis.}
\label{fig:low_ntu}
\end{figure}

Within the 103-sample paired device-transfer subset, full-image Huawei
zero-shot performance remained close to the corresponding Samsung
out-of-fold performance throughout the three low-NTU ranges
(Figure~\ref{fig:low_ntu}A). Below 10 NTU, Samsung produced one
false-negative and one false-positive error, whereas Huawei zero-shot
prediction produced two false-negative errors and one false-positive error.
Cross-fitted log-affine calibration eliminated false-negative errors in this
subset but introduced three false-positive errors and increased numerical
error above 1 NTU. Therefore, global post-hoc calibration did not provide a
uniformly preferable low-NTU operating point.

The seven samples from 4 to 6 NTU are shown separately in
Figure~\ref{fig:low_ntu}C because even a
small numerical error in this interval can move a prediction across the 5-NTU
decision threshold. This panel directly compares reference and predicted NTU
for both devices and for the exploratory calibrated second-device result; it
is descriptive because the subset is too small for a standalone performance
claim.

\subsection{Repeatability near 1.87 NTU}\label{subsec:repeatability_results}

\begin{figure}[t]
\centering
% ASSET CORRECTION REQUIRED: Update artwork label to P(reference turbidity <5 NTU); use ConvNeXt-Tiny consistently.
\includegraphics[width=\textwidth]{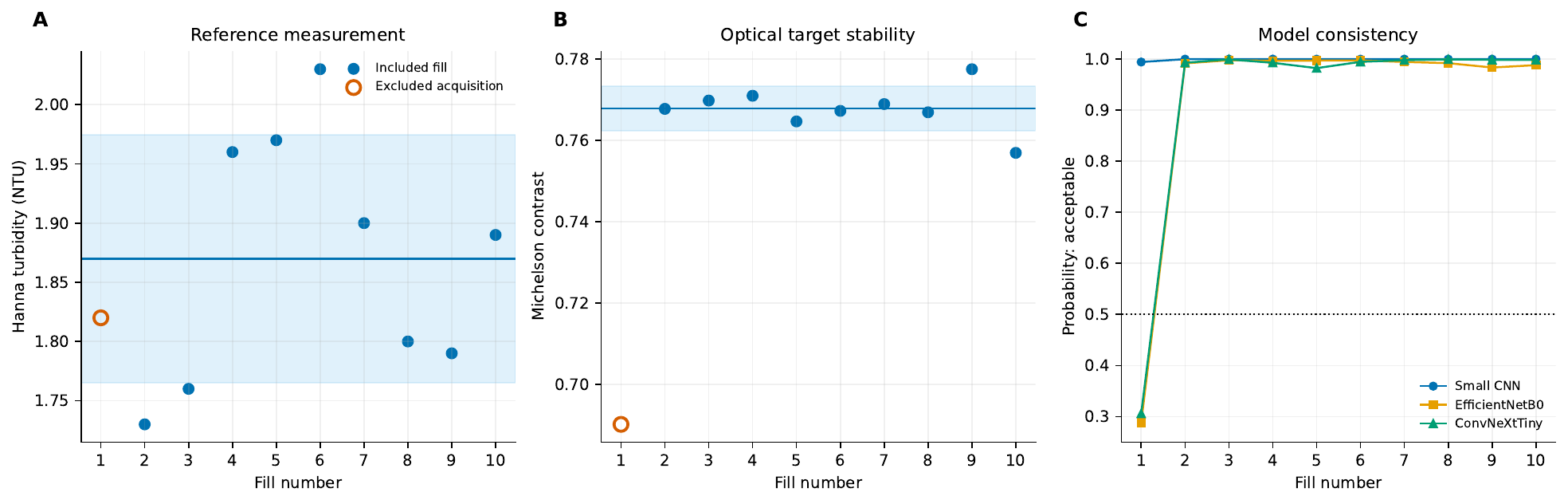}
\caption{Repeat-fill stability near 1.87 NTU, with the excluded
acquisition shown for transparency. \textbf{A}, Hanna reference turbidity across
ten fills. \textbf{B}, Michelson contrast of the optical target.
\textbf{C}, Predicted probability of the below-5-NTU category from the three
classification models. Filled markers represent the nine valid fills in the
primary analysis; the open orange marker identifies the first acquisition,
which was excluded based on independently confirmed acquisition or labeling
provenance rather than model output. Horizontal lines and shaded bands in
panels A and B show the primary-analysis mean and mean $\pm$ one SD. This
experiment characterizes the combined refill, reference-measurement, imaging,
and prediction procedure at one tested low-turbidity condition, not full-range
measurement uncertainty.}
\label{fig:repeatability}
\end{figure}

For the nine valid repeated fills from the same bulk sample, the reference turbidimeter mean was 1.870 NTU,
with a standard deviation (SD) of 0.105 NTU, a coefficient of variation of
5.61\%, and a range of 1.73--2.03 NTU.
Across the tested optical features, coefficients of variation ranged from 0.63\% to 2.67\%.
All three classifiers correctly assigned all nine primary fills to the
below-5-NTU category. Figure~\ref{fig:repeatability} shows the reference
measurements, target contrast, and classifier outputs across the repeated fills.

When all ten observations were retained as a sensitivity analysis, the
reference-measurement summary changed little. In contrast, the excluded
acquisition showed a pronounced shift in several optical features and caused
EfficientNetB0 and ConvNeXt-Tiny to predict the at-or-above-5-NTU category.
The small CNN retained the below-5-NTU prediction. Because exclusion was based
on independently confirmed acquisition provenance rather than model behavior,
this sensitivity result illustrates the importance of acquisition quality
control without changing the nine-fill primary analysis.

\subsection{Direct cross-device transfer without retraining}\label{subsec:device_transfer_results}

\begin{figure}[t]
\centering
\includegraphics[width=\textwidth]{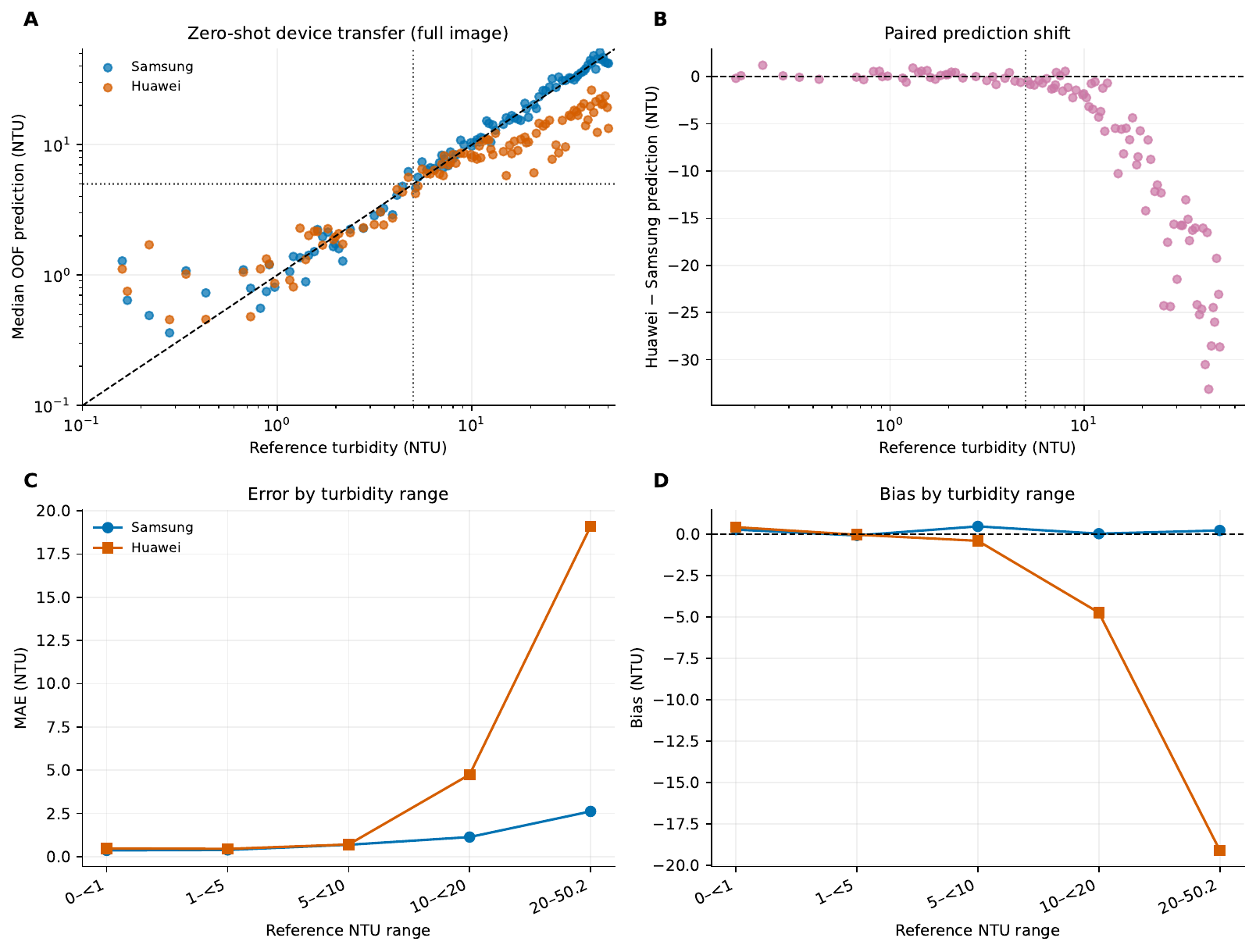}
\caption{Full-image zero-shot transfer of the Samsung-trained EfficientNetB0
regression model to paired Huawei images. \textbf{A}, Reference turbidity
versus median repeated out-of-fold prediction for Samsung and Huawei images of
the same 103 physical samples. The dashed identity line and dotted 5-NTU
threshold lines are shown for reference. \textbf{B}, Paired
Huawei-minus-Samsung prediction difference versus reference turbidity.
\textbf{C}, Device-specific MAE by predefined NTU range. \textbf{D},
Device-specific bias, defined as prediction minus reference NTU, by range.
Huawei images were not used for neural-network training, early stopping,
preprocessing selection, or original model selection. Range-specific values
in panels C and D are descriptive point estimates.}
\label{fig:device_transfer}
\end{figure}

For the full-image representation, Huawei zero-shot MAE remained close to the
Samsung in-domain value at low turbidity: 0.474 versus 0.360 NTU below 1 NTU,
0.448 versus 0.385 NTU from 1 to below 5 NTU, and 0.701 versus 0.684 NTU from
5 to below 10 NTU. Device-dependent divergence emerged at higher turbidity.
In Figure~\ref{fig:device_transfer}A the logarithmic axes visually enlarge
small absolute offsets near zero; the corresponding range-specific MAE values
above provide the more interpretable NTU-scale comparison for that region.
From 10 to below 20 NTU, Huawei MAE was 4.743 NTU compared with 1.131 NTU for
Samsung; from 20 to 50.2 NTU, the corresponding values were 19.095 and
2.613 NTU. Huawei bias in these ranges was $-4.743$ and $-19.095$ NTU,
respectively, demonstrating progressively stronger underestimation.

The target-crop representation exhibited substantially poorer zero-shot
Huawei transfer despite its similar Samsung in-domain classification
performance. This contrast indicates that similar in-domain spatial-ablation performance
does not by itself establish robustness across devices. Because the
full-image preference was informed by the paired Huawei results, it is
reported as an observed representation-dependent transfer result rather than
as a prospectively selected configuration.

A separate post-hoc sensitivity analysis evaluated raw and orientation-matched
Huawei inputs using identical fitted models within each comparison. For the
full-image representation, rotating the Huawei images by $90^\circ$
counterclockwise reduced MAE from 6.978 to 6.457 NTU and RMSE from 11.522 to
10.373 NTU. The paired MAE reduction was 0.521 NTU (95\% stratified
physical-sample bootstrap interval 0.287--0.755), but substantial
high-turbidity underestimation remained.

For the target-crop representation, orientation matching reduced MAE only
from 10.137 to 9.941 NTU and changed RMSE from 14.247 to 14.177 NTU. The
paired reduction in MAE was 0.195 NTU (95\% interval 0.040--0.359), whereas
the RMSE difference was inconclusive. Threshold accuracy decreased from
0.738 to 0.728, with false-negative errors increasing from 27 to 28.
Orientation matching therefore did not explain the representation-dependent
transfer failure.

Because the orientation sensitivity analyses used separately fitted repeated
cross-validation runs, their raw baselines differ slightly from the primary
device-transfer estimates. Orientation effects were interpreted only through
the paired within-run comparisons. Complete results are reported in
Supplementary Table S8.

\subsection{Exploratory post-hoc Huawei calibration}\label{subsec:calibration_results}

For the full-image representation, cross-fitted log-affine calibration reduced Huawei MAE
from 6.942 to 3.323 NTU and RMSE from 11.419 to 5.910 NTU, while changing bias from
$-6.700$ to $-0.816$ NTU. For the target crop, it reduced MAE from 10.044 to 3.392 NTU
and RMSE from 14.107 to 5.708 NTU. These aggregate improvements were concentrated at higher
NTU; full-image calibration increased MAE from 0.448 to 0.675 NTU in the 1--$<$5 range and
from 0.701 to 2.559 NTU in the 5--$<$10 range.

These results motivate prospective range-aware or decision-aware calibration
rather than a hybrid calibration rule defined after inspecting the present
data. The corresponding measured-versus-predicted and residual distributions
are shown in Supplementary Figure~S1. Complete results for the multiplier,
log-affine, and isotonic calibrators, together with calibration-parameter
stability, are reported in the Supplementary Information.

% ============================================================================
\section{Discussion}
\label{sec:discussion}

\subsection{Principal findings}

This study evaluated controlled-illumination smartphone imaging as a method
for estimating continuous turbidity and supporting threshold-oriented
screening. In the primary Samsung analysis, repeated grouped
cross-validation produced an MAE of 1.208 NTU and an $R^2$ of 0.9816 across
162 physical samples. Thus, the average absolute difference from the
reference measurement was about 1.2 NTU across the full range and the predictions captured
more than 98\% of the observed NTU variance, representing strong internal
performance under the controlled laboratory protocol. Numerical error was substantially smaller at low
turbidity, with MAE values of 0.335 NTU below 1 NTU, 0.449 NTU from 1 to
below 5 NTU, and 0.718 NTU from 5 to below 10 NTU. When the continuous
predictions were interpreted relative to the 5-NTU operational threshold,
accuracy was 0.9753, although threshold-side errors remained.

The repeatability experiment showed stable optical features and consistent
below-5-NTU classification across nine valid fills from one bulk suspension
near 1.87 NTU. This finding supports short-term procedural stability under
the tested condition but does not establish repeatability across the complete
turbidity range. It cannot be extrapolated quantitatively to other NTU values;
future repeatability studies should include independent fills under near-clear-water,
sub-1-NTU, 5-NTU boundary, intermediate, and high-turbidity conditions.

Paired-device differences were anticipated because the two camera pipelines
were not expected to be interchangeable. The notable result was that direct
full-image transfer remained close to the development-device result below
10 NTU, despite the absence of retraining or calibration. As anticipated,
larger differences emerged at higher turbidity, where Huawei predictions
showed progressively stronger negative bias. The target-crop representation
transferred considerably less effectively, despite having modestly higher
registered Samsung classification point estimates. Correcting the
identified pixel-orientation mismatch produced only modest improvements and
did not eliminate the high-NTU underestimation. Cross-fitted post-hoc
calibration reduced aggregate Huawei error, particularly at higher turbidity,
but degraded numerical performance in parts of the low-NTU range.

\subsection{Why spatial representation matters}

The registered spatial analyses showed that representation performance was
task-dependent. The central target crop produced the highest observed
binary-classification point estimates, although its advantage over the full
image was modest and the paired intervals included zero. In contrast, the full
image produced the lowest observed aggregate continuous-regression error, and
the paired regression comparison did not establish a clear difference between
the full image and target crop. Restricting the input to twelve light
neighborhoods increased continuous-regression error relative to the full
image. Together with the poorer target-crop Huawei transfer, these results
indicate that strong in-domain threshold discrimination does not by itself
identify the representation most useful for numerical estimation or
cross-device robustness.

One possible explanation is that the wider acquisition image contains
contextual information related to illumination, enclosure geometry, and
global intensity response that helps stabilize predictions when the camera
pipeline changes. Conversely, a restricted crop may emphasize local
colorimetric or contrast characteristics that are more sensitive to differences
in sensor response, optics, exposure, or image processing. These mechanisms
remain hypotheses because the present study was not designed to isolate the
individual contribution of each image region or camera-processing stage.

The orientation sensitivity analysis provides an additional qualification.
Aligning Huawei pixels with the stored Samsung orientation improved some
aggregate metrics, particularly for the full-image representation, but the
principal device-dependent bias remained. Pixel orientation was therefore a
contributing preprocessing difference rather than the main explanation for
the observed domain shift.

\subsection{Low-NTU screening and decision risk}

The relatively small errors below 10 NTU are encouraging for applications
focused on low-turbidity screening. Turbidity is a principal operational
indicator of water clarity and treatment performance, and the present system
could expand where and how frequently it is assessed. The 5-NTU threshold
used here is an operational decision point for evaluating model behavior; its
interpretation in practice should follow the intended monitoring context and
applicable standards.

For screening applications, false-negative and false-positive errors at the
5-NTU threshold have different consequences. A false-negative result fails
to flag a sample whose reference turbidity is at or above the operational
threshold, whereas a false-positive result flags a sample below that
threshold and may increase unnecessary follow-up. Neither error category
alone establishes whether water is safe or requires treatment. The appropriate
operating threshold should therefore depend on the intended use and relative
error costs. For a safety-oriented workflow, it may be preferable to select
a prediction threshold that prioritizes sensitivity rather than applying a
fixed 5-NTU prediction cutoff.

The high area under the receiver operating characteristic curve (ROC AUC)
and average precision derived from the Samsung regression
indicate strong ranking discrimination, but these measures do not guarantee
accurate numerical estimation or transfer to another device. Moreover, the
4--6 NTU boundary subset contained only seven paired samples. Conclusions
about performance immediately around the operational threshold must therefore
remain descriptive until denser boundary-region data are collected.

\subsection{Repeatability, acquisition quality control, and uncertainty}

The repeatability experiment assessed the combined refill,
reference-measurement, settling, imaging, feature-extraction, and prediction
procedure at one low-turbidity condition. Across the nine valid fills, the
reference instrument had a coefficient of variation of 5.61\%, while the
selected optical features had coefficients of variation between 0.63\% and
2.67\%. All three classification models assigned the nine fills to the
below-5-NTU category.

The acquisition excluded because of the independently confirmed acquisition
or labeling error provides an important quality-control example. Although including it changed the reference-turbidity summary only
slightly, it produced marked shifts in several optical features and changed
the decisions of EfficientNetB0 and ConvNeXt-Tiny. Because exclusion was based
on independently confirmed acquisition provenance rather than model output,
this observation does not alter the primary analysis. Instead, it indicates
that a deployable system should include acquisition-quality checks for
unexpected geometry, labeling, illumination, or sample-handling conditions.

These findings should not be interpreted as a full measurement-uncertainty
budget. Repeated acquisitions are still required at multiple NTU levels,
across independently prepared samples, sessions, operators, and devices.
Such experiments would separate reference-instrument variability, sample
preparation, refill and settling effects, imaging variability, and model
variability.

\subsection{Device transfer and the role of calibration}

Controlled illumination and fixed geometry reduce important sources of
variation, but they do not make smartphone cameras interchangeable. Sensors,
optics, exposure control, white balance, tone mapping, compression, and
device-specific image-processing pipelines can all alter the relationship
between the recorded image and turbidity. The Huawei results represent one
handset and its observed acquisition pipeline, including two software builds
and limited acquisition-setting variation; they should not be generalized to
all Huawei devices or smartphones.

The increasing negative Huawei bias at higher turbidity suggests a
device-dependent response compression or scaling difference, although the
present experiment cannot identify its physical or computational cause.
Cross-fitted log-affine calibration partially corrected the aggregate bias and
reduced MAE and RMSE. However, the same global mapping increased error in parts
of the low-NTU range. A calibration optimized only for aggregate full-range
error may therefore be unsuitable for a screening system whose principal risk
lies close to a low operational threshold.

Future calibration should be fitted on a separate deployment dataset and
evaluated prospectively. Potential approaches include standardized internal
reference regions, device-specific intensity normalization, range-aware
calibration, and decision-aware calibration optimized around the intended
screening threshold. An application that standardizes camera responses is a possible
longer-term approach: it could enforce orientation and exposure checks, use
internal color or intensity references, and apply a validated device-response
transformation before turbidity prediction. The objective would be to make
results less dependent on phone manufacturer and model, but that objective
would require validation across multiple phone models, multiple physical
units of each model, software versions, and acquisition sessions.

\subsection{Limitations}

\begin{itemize}
  \item Laboratory milk suspensions may not reproduce the particle-size
  distribution, color, dissolved constituents, or optical behavior of natural
  and treated waters.

  \item The primary Samsung analysis used internal repeated grouped
  cross-validation rather than an independent validation
  dataset from a new population of water samples.

  \item The development manifest contained one image per physical sample, so
  the primary regression analysis could not directly estimate capture
  variability at a fixed turbidity.

  \item The repeatability experiment used one bulk suspension near 1.87 NTU
  and does not characterize repeatability across the full NTU range or
  repeatability of the continuous regression output in NTU.

  \item The device-transfer study used one Huawei handset and the same paired
  physical-sample population. Huawei acquisitions included two software
  builds, ISO values of 50 and 100, and limited variation in aperture,
  exposure time, and white-balance metadata. Transfer behavior therefore
  reflects the combined device and acquisition pipeline and cannot be
  attributed exclusively to camera hardware.

  \item The original neural-network loader did not apply EXIF orientation
  metadata. Orientation normalization was evaluated retrospectively and
  therefore remains a post-hoc sensitivity analysis.

  \item Calibration approaches were developed after inspection of the
  zero-shot device-transfer behavior. Cross-fitting prevented direct
  sample-level leakage but did not provide an independent prospective
  calibration study.

  \item The 4--6 NTU boundary subset contained only seven paired physical
  samples.
\end{itemize}

\subsection{Future work}

The next priority is prospective validation using independently collected
natural and treated-water samples, with denser sampling below 5 NTU and around
the intended decision threshold. Repeated captures should be obtained from
independently prepared samples over multiple sessions and days to quantify
sample-preparation, acquisition, and model variability.

Multi-device evaluation should include several phone models and more than one
physical unit per model. Images should be normalized to a predefined apparatus
orientation before model fitting, and acquisition-quality checks should detect
unexpected geometry, illumination, saturation, or metadata. Device calibration
should be fitted on a separate deployment dataset and assessed both across the
full NTU range and at the intended operational threshold. This work should
support a camera-standardization layer that maps different phone acquisition
pipelines to a common, validated representation before model inference.

If these prospective studies confirm adequate reliability, the system could
support a rapid screening layer in which measurements and associated metadata
are transmitted to a centralized service and displayed in georeferenced
monitoring dashboards. Such deployment should complement validated sampling procedures and
confirmatory laboratory or regulatory measurements.

% ============================================================================
\section{Conclusions}\label{sec:conclusions}

The study demonstrates that controlled-illumination smartphone images can
support continuous turbidity estimation with low error under internal repeated
grouped validation, including strong numerical performance in the low-NTU
range. A post-hoc geometry-normalization sensitivity analysis preserved strong
full-image regression performance and did not demonstrate a continuous-NTU
advantage for the target crop, despite its higher observed binary-classification
point estimates. Short-term repeat-fill results were stable under one tested
low-turbidity condition. Device-dependent differences were expected; notably,
direct full-image transfer to a second smartphone remained comparatively
accurate below 10 NTU without retraining, although substantial
underestimation emerged at higher turbidity and was more severe for the
target-crop representation. Orientation matching did not explain this domain
shift, and post-hoc calibration reduced aggregate error at the cost of poorer
performance in parts of the low-NTU range. These findings motivate a future
camera-standardization layer and prospective multi-device, multi-session, and
natural-water validation before broader deployment.

% ============================================================================
\backmatter

\bmhead{Supplementary information}

The Supplementary Information contains the earlier Samsung architecture
comparison, complete spatial-ablation results, a registered
continuous-regression sensitivity analysis, operational discrimination curves
derived from the primary Samsung regression, target-crop
device-transfer results, cross-fitted Huawei calibration analyses and
parameter stability, the all-ten-observation repeatability sensitivity
analysis, individual predictions for the 4--6 NTU boundary subset, and the
post-hoc input-orientation sensitivity analysis.

\bmhead{Acknowledgements}

The authors thank Francisca Villanueva for the initial suggestion that
motivated this project and for valuable discussions during its early
conceptual development. OpenAI ChatGPT (GPT-5) was used to assist with the organization of the manuscript and with drafting, editing, and improving the clarity of portions of the text. All AI-assisted content was critically reviewed, verified, and revised by the authors, who take full responsibility for the accuracy, integrity, and originality of the manuscript.

\bmhead{Author contributions}

S.O. led the development of the study concept, reviewed the relevant literature, designed
the machine-learning methodology, developed the software and analysis
framework, performed the formal analyses and calibration studies, prepared
the visualizations, supervised and administered the project, acquired funding,
and wrote the original manuscript.

A.C. led the experimental implementation and laboratory work. He developed
and standardized the principal sample-preparation, reference-measurement, and
image-acquisition procedures; designed and fabricated the 3D-printed
acquisition enclosure; and acquired and curated the Samsung and Huawei image
datasets.

P.V. contributed substantially to the conception and experimental methodology
of the study. She initially provided access to laboratory facilities,
co-developed the procedures, quality-control rules, and standardized steps
used for sample preparation and data acquisition, and contributed
physicochemical expertise to the validation and interpretation of the
experimental results. She also acquired funding and critically reviewed and
revised the manuscript.

R.A. provided access to the reference turbidimeter and associated laboratory
resources, instructed the research team in the correct operation of the
instrument, contributed to the development and validation of the measurement
procedure, interpreted the results from a chemistry perspective, and
critically reviewed and revised the manuscript.

N.R. designed and constructed the initial laboratory prototype, performed the
preliminary experiments, acquired the initial dataset, and developed the first
procedure for preparing and imaging milk-based turbidity samples.

All authors reviewed and approved the final manuscript.

\bmhead{Funding}

This work was supported by the Agencia Nacional de Investigaci\'on y Desarrollo
(ANID), Chile, through FONDECYT grant No. 11251121 awarded to S. Olivares;
Project ANID ANILLO ATE 220024; and the Millennium Institute on Green Ammonia
as Energy Vector (MIGA), ANID/Millennium Science Initiative Program,
ICN2021\_023.

\bmhead{Competing interests}
The authors declare no competing interests.

\bmhead{Data availability}

The image dataset, sample metadata, device-pairing manifest, cross-validation
assignments, and source data underlying the figures and tables are available
from the corresponding author to editors, reviewers, and readers on
reasonable request, without undue restriction or delay. A public archival
repository record will be added upon acceptance. Nonessential identifying
metadata will be removed from shared image files. No human-participant or
personal data were collected.

\bmhead{Code availability}

The code used for image preprocessing, model training, repeated grouped
cross-validation, device-transfer evaluation, calibration, statistical
analysis, and generation of the publication figures and tables is available
from the corresponding author to editors, reviewers, and readers on
reasonable request, without undue restriction or delay. A versioned public code
release and archival record will be added upon acceptance, together with the
configuration and environment files required to reproduce the analyses.

% AUTHOR CHECK: Replace planned repository releases with real DOI/URL records
% when available. If data/code remain nonpublic at publication, state the actual
% reason; no reason or repository identifier has been invented in this revision.
%\bibliography{sn-bibliography,submission-review-references}
\bibliography{sn-bibliography}

\end{document}

\maketitle

\section*{Overview}

This Supplementary Information provides additional analyses supporting the
main manuscript, including post-hoc Huawei calibration, target-crop device
transfer, Samsung architecture and spatial-representation comparisons,
operational threshold-discrimination curves, calibration details,
repeatability sensitivity analyses, sample-level results near the 5-NTU
operational threshold, and the post-hoc input-orientation sensitivity
analysis.

% ============================================================================
\section{Cross-fitted post-hoc Huawei calibration}

The uncalibrated Huawei evaluation reported in the main manuscript remains the
primary device-transfer result. Calibration was specified after inspection of
the zero-shot device-transfer behavior and is therefore considered exploratory.
All calibration predictions were obtained by cross-fitting so that a physical
sample's reference NTU was not used to fit the calibrator applied to that
sample.

\begin{figure}[ht]
\centering
\includegraphics[width=\textwidth]
{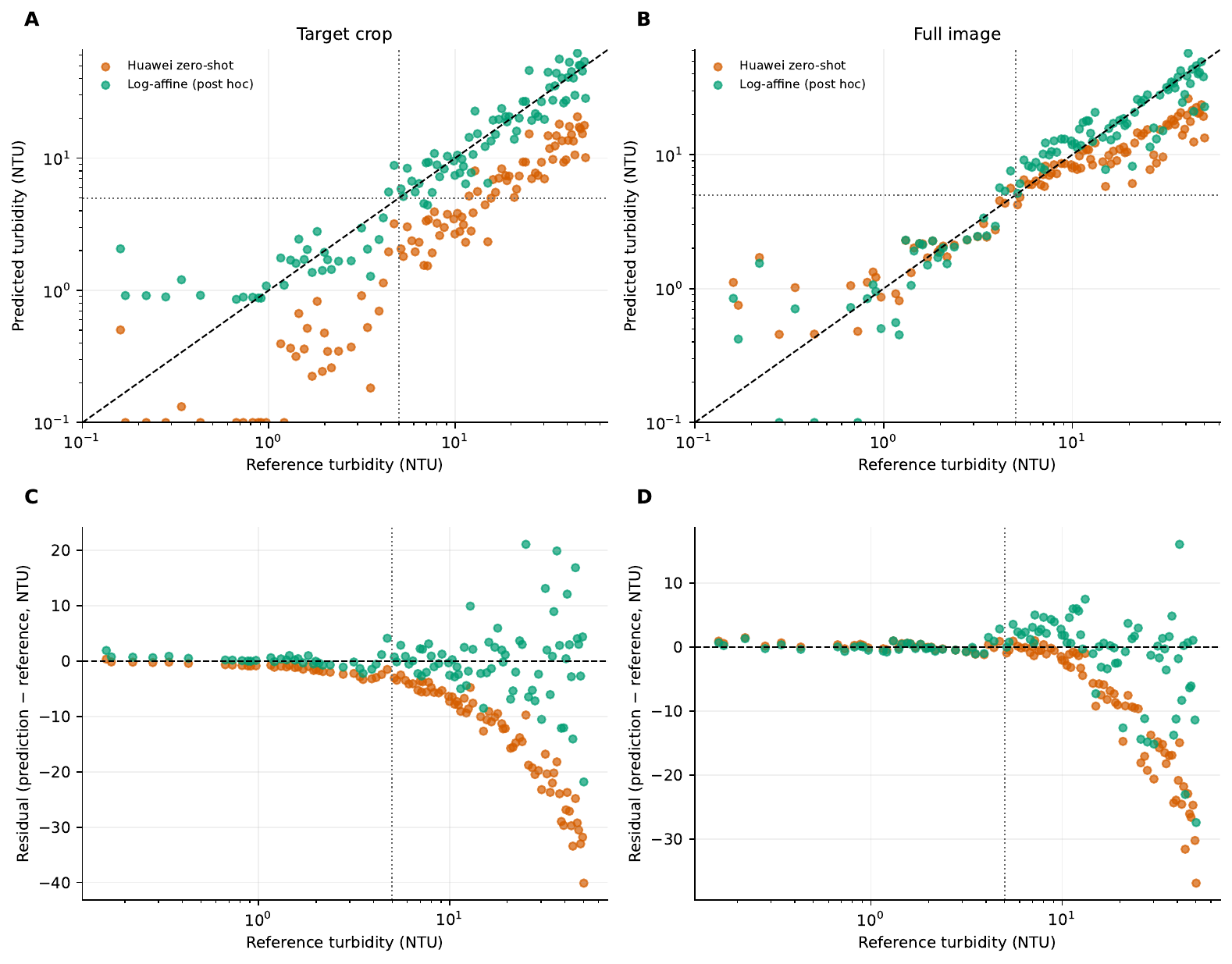}
\caption{\textbf{Post-hoc cross-fitted Huawei calibration.}
Measured-versus-predicted relationships and residuals for the target-crop
(\textbf{A}, \textbf{C}) and full-image (\textbf{B}, \textbf{D})
representations. Orange markers show zero-shot Huawei predictions and green
markers show cross-fitted log-affine predictions. Every physical sample was
evaluated only using calibrators fitted without that sample. Cross-fitting
prevents direct sample leakage but does not constitute independent prospective
device calibration.}
\label{fig:supp_calibration}
\end{figure}
\FloatBarrier

% ============================================================================
\section{Target-crop device-transfer results}

The target-crop representation was evaluated using the same 103 paired
physical samples and repeated cross-validation assignments as the full-image
device-transfer analysis. For each representation, the corresponding
Samsung-trained regression models were applied to the paired held-out
Samsung and Huawei images.
% AUTHOR CHECK: Confirm from the training records that each representation
% used its own fitted weights; this wording does not assert shared weights.
Although the target crop produced
modestly higher point estimates than the full image in the registered Samsung 
binary-classification analysis, its zero-shot transfer to Huawei images was substantially 
poorer, particularly at higher turbidity.

Huawei target-crop predictions showed progressively stronger underestimation
as reference turbidity increased. This result demonstrates that similar
in-domain performance across spatial representations does not necessarily
imply equivalent cross-device robustness. Because the representation comparison
was examined using the paired Huawei results, it is interpreted as an observed
representation-dependent transfer result rather than a prospectively selected
configuration.

\begin{figure}[htbp]
\centering
\includegraphics[width=\textwidth]
{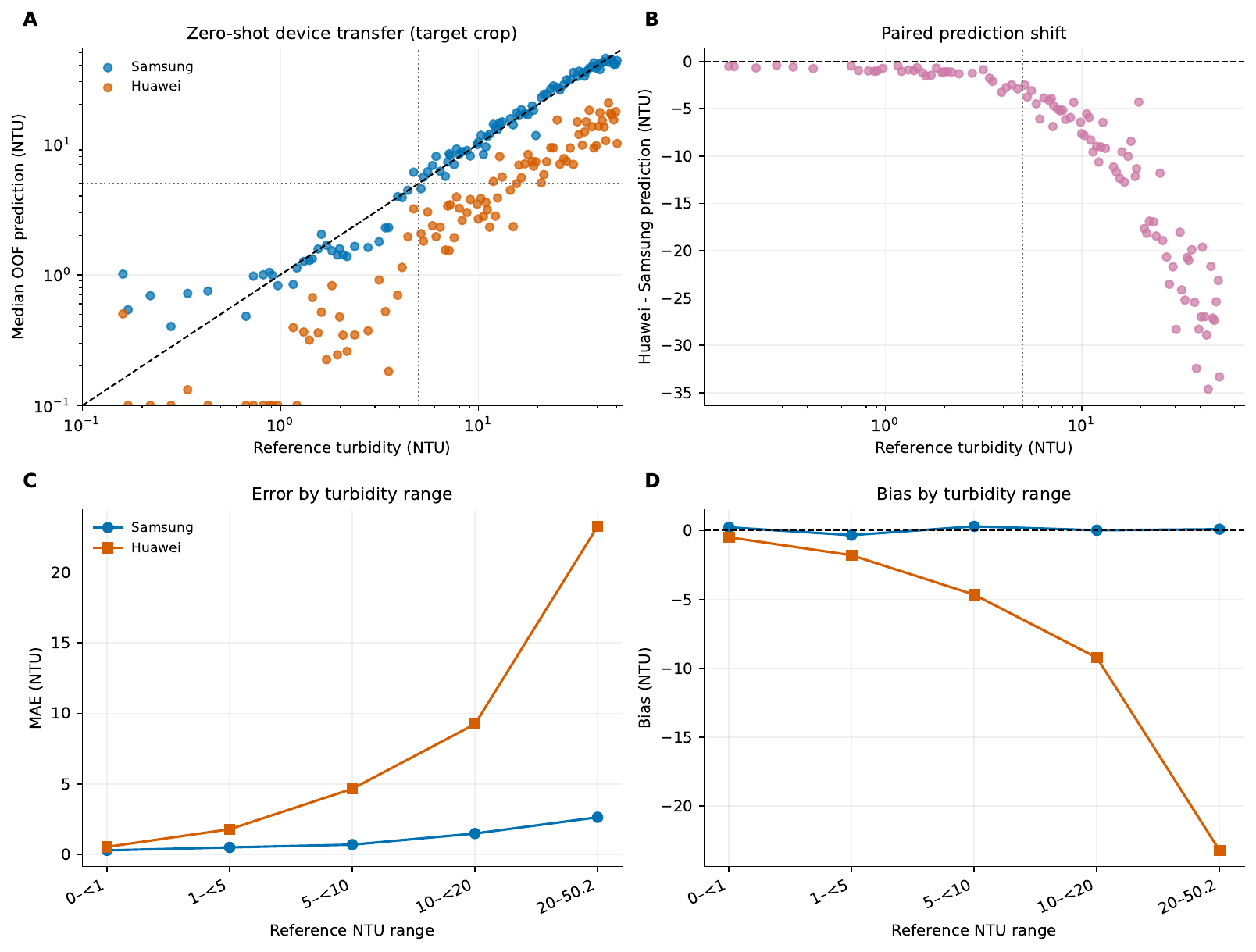}
\caption{\textbf{Target-crop zero-shot transfer of the Samsung-trained
EfficientNetB0 regression model to paired Huawei images.}
\textbf{A}, Reference turbidity versus median repeated out-of-fold prediction
for Samsung and Huawei target-crop images of the same 103 physical samples.
The dashed line represents identity.
\textbf{B}, Paired Huawei-minus-Samsung prediction difference versus reference
turbidity.
\textbf{C}, Device-specific mean absolute error (MAE) by predefined turbidity
range.
\textbf{D}, Device-specific bias, defined as prediction minus reference NTU,
by range. Huawei images were not used for neural-network training, early
stopping, preprocessing selection, or original model selection. For the
logarithmic display in panel A, predictions below 0.1 NTU are shown at the
0.1-NTU plotting floor. Range-specific results are descriptive point
estimates.}
\label{fig:supp_target_crop_transfer}
\end{figure}
\FloatBarrier

% ============================================================================
\section{Samsung architecture comparison}

The small CNN, EfficientNetB0, and ConvNeXt-Tiny were compared in the earlier
Samsung binary-classification experiment using a held-out test set of 25
images. These results provide supporting evidence for the selection of
EfficientNetB0, but they are not part of the repeated grouped
continuous-NTU regression evaluation and do not assess Huawei device transfer.
Discrimination is summarized numerically using ROC AUC and average precision
in Table~\ref{tab:supp_architecture_metrics}. Because this preliminary
comparison used only one held-out split, separate ROC, precision--recall, and
training-history figures are not included.

% ASSET CORRECTION REQUIRED: Display Small CNN, EfficientNetB0, and
% ConvNeXt-Tiny in this external table; verify AP versus PR AUC naming.
\begin{longtable}{@{}lrrrrrrr@{}}
\caption{Samsung-only architecture results from the original held-out classification experiment.}\label{tab:supp_architecture_metrics}\\
\toprule
Architecture & Accuracy & Balanced accuracy & Macro F1 & ROC AUC & PR AUC & MCC & Kappa \\
\midrule
\endfirsthead
\caption[]{Samsung-only architecture results from the original held-out classification experiment. (continued)}\\
\toprule
Architecture & Accuracy & Balanced accuracy & Macro F1 & ROC AUC & PR AUC & MCC & Kappa \\
\midrule
\endhead
\midrule
\multicolumn{8}{r}{Continued on next page}\\
\endfoot
\bottomrule
\endlastfoot
small\_cnn & 0.880 & 0.882 & 0.873 & 0.951 & 0.979 & 0.748 & 0.746 \\
efficientnetb0 & 0.960 & 0.969 & 0.958 & 1.000 & 1.000 & 0.919 & 0.915 \\
convnext\_tiny & 0.920 & 0.938 & 0.917 & 1.000 & 1.000 & 0.846 & 0.834 \\
\end{longtable}

% ============================================================================
\section{Complete spatial-ablation results}

% SUBMISSION BLOCKER: The supplied PDF's Table S2 does not match the registered
% classification results in main Figure 3 and Section 4.2. Regenerate this
% external table from the SAME registered predictions used for Figure 3.
% Do not infer numerical table entries from the plotted points or replace them
% with the continuous-regression results in Table S3.
% Clearly distinguish fold mean/SD from sample-level aggregated paired metrics.
\begin{longtable}{@{}llrrr@{}}
\caption{Complete Samsung spatial-ablation results. Fold SD is descriptive; paired intervals use stratified physical-sample bootstrap resampling conditional on the fitted models.}\label{tab:supp_ablation}\\
\toprule
Representation/comparison & Metric & Mean & SD or difference & 95\% interval \\
\midrule
\endfirsthead
\caption[]{Complete Samsung spatial-ablation results. Fold SD is descriptive; paired intervals use stratified physical-sample bootstrap resampling conditional on the fitted models. (continued)}\\
\toprule
Representation/comparison & Metric & Mean & SD or difference & 95\% interval \\
\midrule
\endhead
\midrule
\multicolumn{5}{r}{Continued on next page}\\
\endfoot
\bottomrule
\endlastfoot
Full image & Accuracy & 0.9763 & 0.0242 & -- \\
Full image & Balanced accuracy & 0.9732 & 0.0271 & -- \\
Full image & Macro F1 & 0.9743 & 0.0262 & -- \\
Full image & MCC & 0.9495 & 0.0521 & -- \\
Full image & Cohen's kappa & 0.9487 & 0.0523 & -- \\
Full image & ROC AUC & 0.9973 & 0.0036 & -- \\
Central target crop & Accuracy & 0.9677 & 0.0247 & -- \\
Central target crop & Balanced accuracy & 0.9618 & 0.0266 & -- \\
Central target crop & Macro F1 & 0.9648 & 0.0262 & -- \\
Central target crop & MCC & 0.9316 & 0.0514 & -- \\
Central target crop & Cohen's kappa & 0.9297 & 0.0523 & -- \\
Central target crop & ROC AUC & 0.9946 & 0.0050 & -- \\
Twelve light neighborhoods & Accuracy & 0.9605 & 0.0189 & -- \\
Twelve light neighborhoods & Balanced accuracy & 0.9506 & 0.0192 & -- \\
Twelve light neighborhoods & Macro F1 & 0.9565 & 0.0203 & -- \\
Twelve light neighborhoods & MCC & 0.9149 & 0.0407 & -- \\
Twelve light neighborhoods & Cohen's kappa & 0.9130 & 0.0407 & -- \\
Twelve light neighborhoods & ROC AUC & 0.9928 & 0.0082 & -- \\
Full Image Minus Target Crop & Accuracy & -- & 0.0000 & [-0.0185, 0.0185] \\
Full Image Minus Target Crop & Balanced Accuracy & -- & 0.0038 & [-0.0144, 0.0259] \\
Full Image Minus Target Crop & Macro F1 & -- & 0.0002 & [-0.0196, 0.0207] \\
Full Image Minus Target Crop & Mcc & -- & 0.0001 & [-0.0390, 0.0399] \\
Full Image Minus Target Crop & Kappa & -- & 0.0004 & [-0.0392, 0.0414] \\
Full Image Minus Target Crop & Roc Auc & -- & 0.0038 & [-0.0005, 0.0108] \\
Full Image Minus Light Neighborhoods & Accuracy & -- & 0.0123 & [-0.0185, 0.0432] \\
Full Image Minus Light Neighborhoods & Balanced Accuracy & -- & 0.0211 & [-0.0096, 0.0555] \\
Full Image Minus Light Neighborhoods & Macro F1 & -- & 0.0141 & [-0.0192, 0.0478] \\
Full Image Minus Light Neighborhoods & Mcc & -- & 0.0267 & [-0.0382, 0.0933] \\
Full Image Minus Light Neighborhoods & Kappa & -- & 0.0281 & [-0.0384, 0.0956] \\
Full Image Minus Light Neighborhoods & Roc Auc & -- & 0.0038 & [-0.0003, 0.0101] \\
Target Crop Minus Light Neighborhoods & Accuracy & -- & 0.0123 & [-0.0123, 0.0370] \\
Target Crop Minus Light Neighborhoods & Balanced Accuracy & -- & 0.0172 & [-0.0058, 0.0479] \\
Target Crop Minus Light Neighborhoods & Macro F1 & -- & 0.0139 & [-0.0130, 0.0418] \\
Target Crop Minus Light Neighborhoods & Mcc & -- & 0.0267 & [-0.0270, 0.0812] \\
Target Crop Minus Light Neighborhoods & Kappa & -- & 0.0277 & [-0.0260, 0.0833] \\
Target Crop Minus Light Neighborhoods & Roc Auc & -- & 0.0000 & [-0.0045, 0.0043] \\
\end{longtable}

\subsection{Registered continuous-NTU regression sensitivity analysis}

Following deterministic geometric registration, continuous-NTU regression
was repeated for the full image, central target crop, and twelve-light-
neighborhood representations using the same 162 physical samples, fold
assignments, repetition seeds, model architecture, and training settings.
Because registration was introduced after visual inspection of the
representation panels, this analysis is treated as a secondary sensitivity
analysis rather than as a replacement for the primary full-image regression.

The full image had the lowest observed MAE and RMSE, although paired intervals
did not establish a clear difference between the full image and target crop.
The twelve-light-neighborhood representation produced higher error than the
full image. Complete estimates and paired comparisons are reported in
Table~\ref{tab:supp_registered_regression}.

\begin{landscape}
\begin{table}[p]
\centering
\scriptsize
\setlength{\tabcolsep}{3.2pt}
\caption{Registered continuous-NTU regression sensitivity analysis. Panel A reports estimates with 95\% stratified physical-sample bootstrap intervals in brackets. Panel B reports paired differences, defined as the left representation minus the right representation, with corresponding 95\% intervals. For MAE, RMSE, and log-scale MAE, a negative difference favors the left representation; for $R^2$ and threshold accuracy, a positive difference favors the left representation. Predictions were the median of five repeated out-of-fold predictions for each of 162 physical samples.}
\label{tab:supp_registered_regression}

\textbf{A. Representation-specific performance}\\[0.4em]
\begin{tabular}{@{}lcccccccc@{}}
\toprule
Representation & MAE & RMSE & Median AE & Bias & $R^2$ & Spearman $\rho$ & Log-scale MAE & Threshold accuracy \\
\midrule
Full image
& \shortstack{1.250\\{[1.090, 1.422]}}
& \shortstack{1.791\\{[1.503, 2.124]}}
& \shortstack{0.891\\{[0.648, 1.077]}}
& \shortstack{0.224\\{[$-0.050$, 0.493]}}
& \shortstack{0.9840\\{[0.9767, 0.9888]}}
& \shortstack{0.9901\\{[0.9852, 0.9928]}}
& \shortstack{0.1219\\{[0.1070, 0.1374]}}
& \shortstack{0.9691\\{[0.9383, 0.9938]}} \\
Central target crop
& \shortstack{1.353\\{[1.163, 1.555]}}
& \shortstack{2.024\\{[1.705, 2.360]}}
& \shortstack{0.793\\{[0.589, 0.956]}}
& \shortstack{0.045\\{[$-0.268$, 0.341]}}
& \shortstack{0.9795\\{[0.9722, 0.9854]}}
& \shortstack{0.9896\\{[0.9846, 0.9923]}}
& \shortstack{0.1255\\{[0.1093, 0.1421]}}
& \shortstack{0.9753\\{[0.9506, 0.9938]}} \\
Twelve light neighborhoods
& \shortstack{1.667\\{[1.462, 1.873]}}
& \shortstack{2.331\\{[2.026, 2.625]}}
& \shortstack{1.051\\{[0.916, 1.374]}}
& \shortstack{0.348\\{[0.001, 0.676]}}
& \shortstack{0.9728\\{[0.9649, 0.9799]}}
& \shortstack{0.9849\\{[0.9781, 0.9888]}}
& \shortstack{0.1494\\{[0.1312, 0.1675]}}
& \shortstack{0.9753\\{[0.9506, 0.9938]}} \\
\bottomrule
\end{tabular}

\vspace{1.2em}
\textbf{B. Paired representation differences}\\[0.4em]
\begin{tabular}{@{}lccccc@{}}
\toprule
Comparison & $\Delta$MAE & $\Delta$RMSE & $\Delta R^2$ & $\Delta$ log-scale MAE & $\Delta$ threshold accuracy \\
\midrule
Target crop minus full image
& \shortstack{0.103\\{[$-0.102$, 0.310]}}
& \shortstack{0.233\\{[$-0.118$, 0.571]}}
& \shortstack{$-0.0045$\\{[$-0.0113$, 0.0024]}}
& \shortstack{0.0036\\{[$-0.0119$, 0.0182]}}
& \shortstack{0.0062\\{[$-0.0123$, 0.0309]}} \\
Light neighborhoods minus full image
& \shortstack{0.416\\{[0.171, 0.666]}}
& \shortstack{0.540\\{[0.136, 0.950]}}
& \shortstack{$-0.0111$\\{[$-0.0198$, $-0.0027$]}}
& \shortstack{0.0276\\{[0.0058, 0.0488]}}
& \shortstack{0.0062\\{[$-0.0123$, 0.0309]}} \\
Target crop minus light neighborhoods
& \shortstack{$-0.313$\\{[$-0.559$, $-0.057$]}}
& \shortstack{$-0.307$\\{[$-0.693$, 0.101]}}
& \shortstack{0.0067\\{[$-0.0023$, 0.0155]}}
& \shortstack{$-0.0239$\\{[$-0.0432$, $-0.0041$]}}
& \shortstack{0.0000\\{[$-0.0185$, 0.0185]}} \\
\bottomrule
\end{tabular}
\end{table}
\end{landscape}

% ============================================================================
\section{Operational discrimination of the primary Samsung regression}

Continuous predictions from the primary Samsung regression were evaluated as
scores for distinguishing samples below 5 NTU from samples at or above 5 NTU.
The analysis used one median repeated out-of-fold prediction for each of the
162 physical samples. The resulting ROC AUC was 0.9975
(95\% bootstrap interval 0.9934--1.0000), and average precision was 0.9986
(0.9964--1.0000).

At the prespecified prediction threshold of 5 NTU, sensitivity for samples at
or above 5 NTU was 0.9808, specificity for samples below 5 NTU was 0.9655,
and precision was 0.9808. The corresponding confusion counts were 102 true
positives, 56 true negatives, two false positives, and two false negatives.
Thus, the model showed strong ranking discrimination, although the fixed
operating threshold did not eliminate threshold-side errors.

\begin{figure}[htbp]
\centering
\includegraphics[width=\textwidth]
{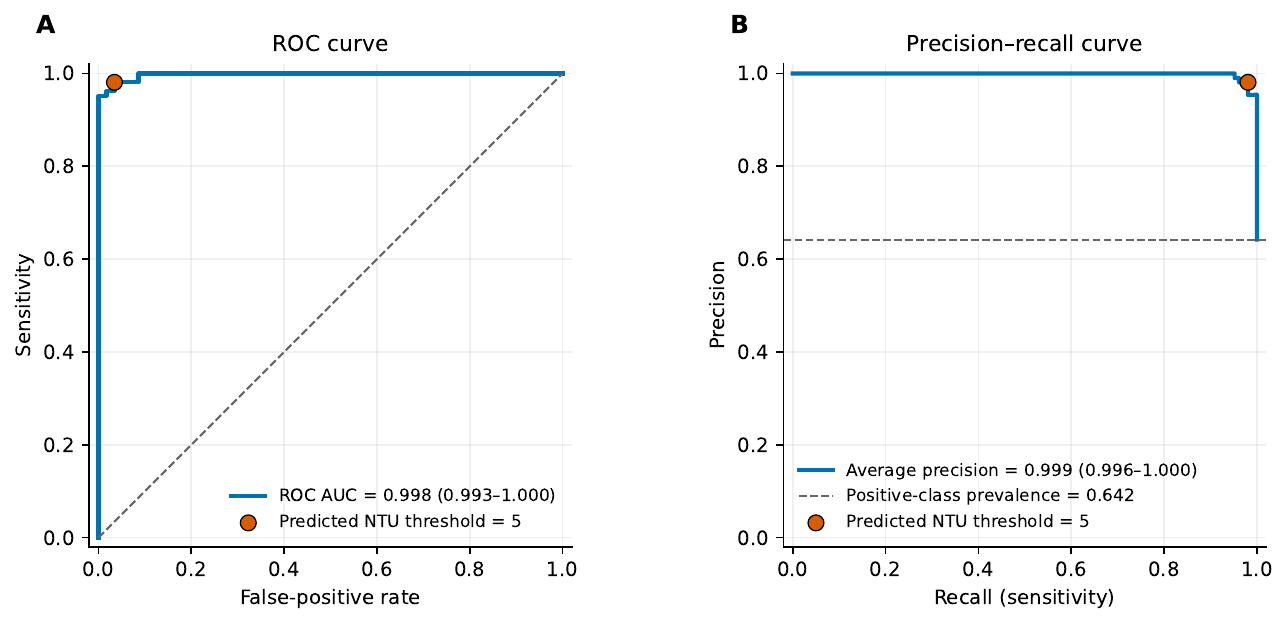}
\caption{\textbf{Operational discrimination derived from the primary Samsung
continuous-NTU regression.}
\textbf{A}, Receiver operating characteristic curve for identifying samples
with reference turbidity at or above 5 NTU using the median repeated
out-of-fold NTU prediction as the decision score.
\textbf{B}, Corresponding precision--recall curve, with samples at or above
5 NTU treated as the positive class. Orange markers show performance when
predicted turbidity is thresholded at 5 NTU. The dashed horizontal line in
panel B represents the positive-class prevalence. Each curve was calculated
using one median out-of-fold prediction for each of 162 physical samples.
The 95\% intervals reported in the legends were obtained using 5000
stratified physical-sample bootstrap repetitions and are conditional on the
fitted repeated cross-validation models and fixed analysis procedure; they
do not incorporate model-training variability.}
\label{fig:supp_samsung_threshold_curves}
\end{figure}

\FloatBarrier

% ============================================================================
\clearpage
\begin{landscape}

\section{Complete calibration results}

\small
\setlength{\tabcolsep}{5pt}
\begin{longtable}{@{}lllrr@{}}
\caption{Complete cross-fitted Huawei calibration results. Calibration is
post hoc and intervals are conditional on the existing neural-network
predictions.}
\label{tab:supp_calibration_methods}\\
\toprule
Representation & Method & Metric & Estimate & 95\% interval \\
\midrule
\endfirsthead

\caption[]{Complete cross-fitted Huawei calibration results. Calibration is
post hoc and intervals are conditional on the existing neural-network
predictions. (continued)}\\
\toprule
Representation & Method & Metric & Estimate & 95\% interval \\
\midrule
\endhead

\midrule
\multicolumn{5}{r}{Continued on next page}\\
\endfoot

\bottomrule
\endlastfoot

Central target crop & Uncalibrated & MAE (NTU) & 10.0443 & [9.2809, 10.7989] \\
Central target crop & Uncalibrated & RMSE (NTU) & 14.1068 & [12.8091, 15.3490] \\
Central target crop & Uncalibrated & Median AE (NTU) & 6.3926 & [5.5886, 7.3121] \\
Central target crop & Uncalibrated & Bias (NTU) & -10.0376 & [-10.7923, -9.2781] \\
Central target crop & Uncalibrated & $R^2$ & 0.0886 & [0.0063, 0.1612] \\
Central target crop & Uncalibrated & Spearman's $\rho$ & 0.9649 & [0.9492, 0.9750] \\
Central target crop & Uncalibrated & Log-scale MAE & 0.8507 & [0.8065, 0.8942] \\
Central target crop & Uncalibrated & Threshold accuracy & 0.7282 & [0.6893, 0.7670] \\
Central target crop & Uncalibrated & Balanced accuracy & 0.7971 & [0.7681, 0.8261] \\
Central target crop & Uncalibrated & Macro-F1 & 0.7269 & [0.6891, 0.7643] \\

Central target crop & Multiplier & MAE (NTU) & 3.1838 & [2.5761, 3.8681] \\
Central target crop & Multiplier & RMSE (NTU) & 5.1939 & [4.0708, 6.3334] \\
Central target crop & Multiplier & Median AE (NTU) & 1.5914 & [1.0824, 2.2133] \\
Central target crop & Multiplier & Bias (NTU) & -0.8491 & [-1.8029, 0.1195] \\
Central target crop & Multiplier & $R^2$ & 0.8765 & [0.8162, 0.9254] \\
Central target crop & Multiplier & Spearman's $\rho$ & 0.9646 & [0.9489, 0.9749] \\
Central target crop & Multiplier & Log-scale MAE & 0.2683 & [0.2324, 0.3073] \\
Central target crop & Multiplier & Threshold accuracy & 0.9515 & [0.9126, 0.9903] \\
Central target crop & Multiplier & Balanced accuracy & 0.9488 & [0.8977, 0.9855] \\
Central target crop & Multiplier & Macro-F1 & 0.9455 & [0.9005, 0.9889] \\

Central target crop & Log-affine & MAE (NTU) & 3.3918 & [2.7351, 4.1339] \\
Central target crop & Log-affine & RMSE (NTU) & 5.7082 & [4.4221, 6.9441] \\
Central target crop & Log-affine & Median AE (NTU) & 1.9761 & [1.2379, 2.2435] \\
Central target crop & Log-affine & Bias (NTU) & 0.0524 & [-1.0344, 1.1519] \\
Central target crop & Log-affine & $R^2$ & 0.8508 & [0.7780, 0.9089] \\
Central target crop & Log-affine & Spearman's $\rho$ & 0.9629 & [0.9472, 0.9735] \\
Central target crop & Log-affine & Log-scale MAE & 0.2218 & [0.1888, 0.2570] \\
Central target crop & Log-affine & Threshold accuracy & 0.9612 & [0.9223, 0.9903] \\
Central target crop & Log-affine & Balanced accuracy & 0.9561 & [0.9118, 0.9928] \\
Central target crop & Log-affine & Macro-F1 & 0.9561 & [0.9122, 0.9891] \\

Central target crop & Isotonic & MAE (NTU) & 3.0644 & [2.5729, 3.5956] \\
Central target crop & Isotonic & RMSE (NTU) & 4.6435 & [3.7438, 5.5648] \\
Central target crop & Isotonic & Median AE (NTU) & 1.7395 & [1.1250, 2.2291] \\
Central target crop & Isotonic & Bias (NTU) & -0.0038 & [-0.8448, 0.8983] \\
Central target crop & Isotonic & $R^2$ & 0.9012 & [0.8493, 0.9367] \\
Central target crop & Isotonic & Spearman's $\rho$ & 0.9536 & [0.9368, 0.9651] \\
Central target crop & Isotonic & Log-scale MAE & 0.2175 & [0.1863, 0.2501] \\
Central target crop & Isotonic & Threshold accuracy & 0.9806 & [0.9515, 1.0000] \\
Central target crop & Isotonic & Balanced accuracy & 0.9706 & [0.9265, 1.0000] \\
Central target crop & Isotonic & Macro-F1 & 0.9777 & [0.9428, 1.0000] \\

Full image & Uncalibrated & MAE (NTU) & 6.9417 & [6.1622, 7.7204] \\
Full image & Uncalibrated & RMSE (NTU) & 11.4188 & [10.0258, 12.7569] \\
Full image & Uncalibrated & Median AE (NTU) & 1.1699 & [0.9782, 1.4866] \\
Full image & Uncalibrated & Bias (NTU) & -6.7003 & [-7.4781, -5.9167] \\
Full image & Uncalibrated & $R^2$ & 0.4028 & [0.3231, 0.4783] \\
Full image & Uncalibrated & Spearman's $\rho$ & 0.9575 & [0.9345, 0.9728] \\
Full image & Uncalibrated & Log-scale MAE & 0.3744 & [0.3389, 0.4098] \\
Full image & Uncalibrated & Threshold accuracy & 0.9709 & [0.9320, 1.0000] \\
Full image & Uncalibrated & Balanced accuracy & 0.9708 & [0.9339, 1.0000] \\
Full image & Uncalibrated & Macro-F1 & 0.9673 & [0.9257, 1.0000] \\

Full image & Multiplier & MAE (NTU) & 4.4519 & [3.8037, 5.1640] \\
Full image & Multiplier & RMSE (NTU) & 6.0408 & [4.9569, 7.2411] \\
Full image & Multiplier & Median AE (NTU) & 3.2260 & [2.9471, 3.9093] \\
Full image & Multiplier & Bias (NTU) & 1.0755 & [0.2058, 1.9194] \\
Full image & Multiplier & $R^2$ & 0.8329 & [0.7686, 0.8856] \\
Full image & Multiplier & Spearman's $\rho$ & 0.9572 & [0.9341, 0.9725] \\
Full image & Multiplier & Log-scale MAE & 0.3774 & [0.3416, 0.4153] \\
Full image & Multiplier & Threshold accuracy & 0.9515 & [0.9126, 0.9806] \\
Full image & Multiplier & Balanced accuracy & 0.9265 & [0.8676, 0.9706] \\
Full image & Multiplier & Macro-F1 & 0.9428 & [0.8931, 0.9777] \\

Full image & Log-affine & MAE (NTU) & 3.3232 & [2.5767, 4.1555] \\
Full image & Log-affine & RMSE (NTU) & 5.9095 & [4.3852, 7.4016] \\
Full image & Log-affine & Median AE (NTU) & 1.3239 & [0.9854, 1.8726] \\
Full image & Log-affine & Bias (NTU) & -0.8163 & [-1.8131, 0.1927] \\
Full image & Log-affine & $R^2$ & 0.8401 & [0.7571, 0.9102] \\
Full image & Log-affine & Spearman's $\rho$ & 0.9567 & [0.9330, 0.9724] \\
Full image & Log-affine & Log-scale MAE & 0.2256 & [0.1894, 0.2648] \\
Full image & Log-affine & Threshold accuracy & 0.9709 & [0.9417, 1.0000] \\
Full image & Log-affine & Balanced accuracy & 0.9559 & [0.9118, 1.0000] \\
Full image & Log-affine & Macro-F1 & 0.9663 & [0.9308, 1.0000] \\

Full image & Isotonic & MAE (NTU) & 3.6709 & [3.0936, 4.3053] \\
Full image & Isotonic & RMSE (NTU) & 5.7231 & [4.6705, 6.7845] \\
Full image & Isotonic & Median AE (NTU) & 1.9563 & [1.4568, 2.3200] \\
Full image & Isotonic & Bias (NTU) & -0.0329 & [-1.0682, 0.9921] \\
Full image & Isotonic & $R^2$ & 0.8500 & [0.7820, 0.9023] \\
Full image & Isotonic & Spearman's $\rho$ & 0.9445 & [0.9181, 0.9618] \\
Full image & Isotonic & Log-scale MAE & 0.2367 & [0.2003, 0.2753] \\
Full image & Isotonic & Threshold accuracy & 0.9709 & [0.9320, 1.0000] \\
Full image & Isotonic & Balanced accuracy & 0.9708 & [0.9339, 1.0000] \\
Full image & Isotonic & Macro-F1 & 0.9673 & [0.9257, 1.0000] \\

\end{longtable}

\begin{table}[htbp]
\centering
\caption{Stability of the cross-fitted log-affine calibration parameters
across five folds and five repeats.}
\label{tab:supp_calibration_parameters}
\small
\begin{tabular}{@{}llrrrrr@{}}
\toprule
Representation & Parameter & $n$ fits & Mean & SD & Minimum & Maximum \\
\midrule
Central target crop & Intercept & 25 & 0.6371 & 0.0216 & 0.5982 & 0.6802 \\
Central target crop & Slope     & 25 & 1.1467 & 0.0151 & 1.1222 & 1.1786 \\
Full image          & Intercept & 25 & $-0.4227$ & 0.0360 & $-0.4831$ & $-0.3450$ \\
Full image          & Slope     & 25 & 1.3550 & 0.0158 & 1.3187 & 1.3795 \\
\bottomrule
\end{tabular}
\end{table}

\end{landscape}
\clearpage

% ============================================================================
\clearpage
\begin{landscape}

\section{Repeatability sensitivity analysis}

\small
\setlength{\tabcolsep}{4pt}
\begin{longtable}{@{}llrrrrrrrr@{}}
\caption{Repeatability summaries for the nine-fill primary analysis and the
all-ten-observation sensitivity analysis.}
\label{tab:supp_repeatability}\\
\toprule
Category & Measure &
$n_9$ & Mean$_9$ & SD$_9$ & CV$_9$ (\%) &
$n_{10}$ & Mean$_{10}$ & SD$_{10}$ & CV$_{10}$ (\%) \\
\midrule
\endfirsthead

\caption[]{Repeatability summaries for the nine-fill primary analysis and the
all-ten-observation sensitivity analysis. (continued)}\\
\toprule
Category & Measure &
$n_9$ & Mean$_9$ & SD$_9$ & CV$_9$ (\%) &
$n_{10}$ & Mean$_{10}$ & SD$_{10}$ & CV$_{10}$ (\%) \\
\midrule
\endhead

\midrule
\multicolumn{10}{r}{Continued on next page}\\
\endfoot

\bottomrule
\endlastfoot

Model output &
ConvNeXt-Tiny $P(<5~\mathrm{NTU})$ &
9 & 0.9951 & 0.0055 & 0.55 &
10 & 0.9262 & 0.2180 & 23.54 \\

Model output &
EfficientNetB0 $P(<5~\mathrm{NTU})$ &
9 & 0.9932 & 0.0049 & 0.49 &
10 & 0.9227 & 0.2229 & 24.16 \\

Model output &
Small CNN $P(<5~\mathrm{NTU})$ &
9 & 0.9999 & 0.0000 & 0.00 &
10 & 0.9993 & 0.0018 & 0.18 \\

Optical feature &
Black-region mean luminance &
9 & 22.4822 & 0.5995 & 2.67 &
10 & 23.4531 & 3.1218 & 13.31 \\

Optical feature &
Michelson contrast &
9 & 0.7678 & 0.0055 & 0.71 &
10 & 0.7601 & 0.0251 & 3.30 \\

Optical feature &
Target saturated fraction &
9 & 0.0246 & 0.0005 & 2.06 &
10 & 0.0247 & 0.0005 & 1.97 \\

Optical feature &
White-to-black luminance ratio &
9 & 7.6183 & 0.2014 & 2.64 &
10 & 7.4020 & 0.7096 & 9.59 \\

Optical feature &
White-region mean luminance &
9 & 171.1705 & 1.0758 & 0.63 &
10 & 171.6170 & 1.7386 & 1.01 \\

Reference measurement &
Hanna turbidity (NTU) &
9 & 1.8700 & 0.1049 & 5.61 &
10 & 1.8650 & 0.1001 & 5.37 \\

\end{longtable}

\end{landscape}
\clearpage

% ============================================================================
\clearpage
\begin{landscape}

\section{Individual 4--6 NTU boundary predictions}

\scriptsize
\setlength{\tabcolsep}{4pt}
\begin{longtable}{@{}lrlllrl@{}}
\caption{Individual predictions for the seven physical samples in the descriptive 4--6 NTU boundary subset.}\label{tab:supp_boundary}\\
\toprule
Sample & Reference NTU & Representation & Configuration & Prediction & Predicted side & Correct \\
\midrule
\endfirsthead
\caption[]{Individual predictions for the seven physical samples in the descriptive 4--6 NTU boundary subset. (continued)}\\
\toprule
Sample & Reference NTU & Representation & Configuration & Prediction & Predicted side & Correct \\
\midrule
\endhead
\midrule
\multicolumn{7}{r}{Continued on next page}\\
\endfoot
\bottomrule
\endlastfoot
4.12 & 4.12 & Full image & Huawei log-affine (post hoc) & 5.658 & At or above 5 & No \\
4.12 & 4.12 & Full image & Huawei zero-shot & 4.532 & Below 5 & Yes \\
4.12 & 4.12 & Full image & Samsung in-domain OOF & 4.094 & Below 5 & Yes \\
4.12 & 4.12 & Central target crop & Huawei log-affine (post hoc) & 3.537 & Below 5 & Yes \\
4.12 & 4.12 & Central target crop & Huawei zero-shot & 1.137 & Below 5 & Yes \\
4.12 & 4.12 & Central target crop & Samsung in-domain OOF & 3.888 & Below 5 & Yes \\
4.4 & 4.40 & Full image & Huawei log-affine (post hoc) & 5.339 & At or above 5 & No \\
4.4 & 4.40 & Full image & Huawei zero-shot & 4.334 & Below 5 & Yes \\
4.4 & 4.40 & Full image & Samsung in-domain OOF & 4.796 & Below 5 & Yes \\
4.4 & 4.40 & Central target crop & Huawei log-affine (post hoc) & 5.547 & At or above 5 & No \\
4.4 & 4.40 & Central target crop & Huawei zero-shot & 1.955 & Below 5 & Yes \\
4.4 & 4.40 & Central target crop & Samsung in-domain OOF & 4.427 & Below 5 & Yes \\
4.71 & 4.71 & Full image & Huawei log-affine (post hoc) & 7.559 & At or above 5 & No \\
4.71 & 4.71 & Full image & Huawei zero-shot & 5.618 & At or above 5 & No \\
4.71 & 4.71 & Full image & Samsung in-domain OOF & 6.213 & At or above 5 & No \\
4.71 & 4.71 & Central target crop & Huawei log-affine (post hoc) & 8.820 & At or above 5 & No \\
4.71 & 4.71 & Central target crop & Huawei zero-shot & 3.187 & Below 5 & Yes \\
4.71 & 4.71 & Central target crop & Samsung in-domain OOF & 6.079 & At or above 5 & No \\
5.13 & 5.13 & Full image & Huawei log-affine (post hoc) & 5.139 & At or above 5 & Yes \\
5.13 & 5.13 & Full image & Huawei zero-shot & 4.223 & Below 5 & No \\
5.13 & 5.13 & Full image & Samsung in-domain OOF & 4.650 & Below 5 & No \\
5.13 & 5.13 & Central target crop & Huawei log-affine (post hoc) & 5.852 & At or above 5 & Yes \\
5.13 & 5.13 & Central target crop & Huawei zero-shot & 2.059 & Below 5 & No \\
5.13 & 5.13 & Central target crop & Samsung in-domain OOF & 4.550 & Below 5 & No \\
5.28 & 5.28 & Full image & Huawei log-affine (post hoc) & 6.077 & At or above 5 & Yes \\
5.28 & 5.28 & Full image & Huawei zero-shot & 4.806 & Below 5 & No \\
5.28 & 5.28 & Full image & Samsung in-domain OOF & 5.621 & At or above 5 & Yes \\
5.28 & 5.28 & Central target crop & Huawei log-affine (post hoc) & 5.127 & At or above 5 & Yes \\
5.28 & 5.28 & Central target crop & Huawei zero-shot & 1.804 & Below 5 & No \\
5.28 & 5.28 & Central target crop & Samsung in-domain OOF & 5.568 & At or above 5 & Yes \\
5.54 & 5.54 & Full image & Huawei log-affine (post hoc) & 9.118 & At or above 5 & Yes \\
5.54 & 5.54 & Full image & Huawei zero-shot & 6.494 & At or above 5 & Yes \\
5.54 & 5.54 & Full image & Samsung in-domain OOF & 7.379 & At or above 5 & Yes \\
5.54 & 5.54 & Central target crop & Huawei log-affine (post hoc) & 8.389 & At or above 5 & Yes \\
5.54 & 5.54 & Central target crop & Huawei zero-shot & 3.023 & Below 5 & No \\
5.54 & 5.54 & Central target crop & Samsung in-domain OOF & 6.123 & At or above 5 & Yes \\
5.86 & 5.86 & Full image & Huawei log-affine (post hoc) & 8.250 & At or above 5 & Yes \\
5.86 & 5.86 & Full image & Huawei zero-shot & 6.016 & At or above 5 & Yes \\
5.86 & 5.86 & Full image & Samsung in-domain OOF & 6.414 & At or above 5 & Yes \\
5.86 & 5.86 & Central target crop & Huawei log-affine (post hoc) & 6.704 & At or above 5 & Yes \\
5.86 & 5.86 & Central target crop & Huawei zero-shot & 2.370 & Below 5 & No \\
5.86 & 5.86 & Central target crop & Samsung in-domain OOF & 6.840 & At or above 5 & Yes \\
\end{longtable}

\end{landscape}
\clearpage

% ============================================================================
\section{Orientation-normalization sensitivity analysis}

The neural-network preprocessing pipeline decoded the stored JPEG pixel arrays
without applying EXIF orientation metadata. All Samsung images carried EXIF
orientation value 6, whereas the Huawei images carried the non-standard value
0. Comparison of fixed apparatus landmarks showed that rotating the Huawei
stored pixels by $90^\circ$ counterclockwise aligned them with the stored
Samsung pixel orientation used for model training.

A post-hoc sensitivity analysis therefore compared the original Huawei inputs
with orientation-matched Huawei inputs. Both versions were processed using the
same crop, resize, and normalization operations and evaluated using identical
fitted Samsung models. Huawei images were not used for neural-network training,
early stopping, or original model selection.

The sensitivity analyses used separately fitted repeated cross-validation
runs. Their raw-input baselines therefore differ slightly from the primary
device-transfer estimates reported elsewhere. Orientation effects were
evaluated only through paired comparisons between raw and rotated inputs
processed by the same fitted models within each sensitivity run.

For the full-image representation, orientation matching reduced Huawei MAE
from 6.978 to 6.457 NTU and RMSE from 11.522 to 10.373 NTU. The paired MAE
reduction was 0.521 NTU (95\% stratified physical-sample bootstrap interval
0.287--0.755). Nevertheless, substantial high-turbidity underestimation
remained, and MAE in the 20--50.2 NTU range was 16.998 NTU after orientation
matching.

For the target-crop representation, orientation matching produced a much
smaller reduction in MAE, from 10.137 to 9.941 NTU. The paired reduction was
0.195 NTU (95\% interval 0.040--0.359), while the RMSE difference was
inconclusive. Moreover, 5-NTU threshold accuracy decreased slightly and the
number of false-negative errors increased from 27 to 28. These findings
indicate that pixel orientation contributed modestly to some numerical errors
but did not account for the principal cross-device domain shift or the poorer
transfer of the target-crop representation.

\begin{table}[htbp]
\centering
\caption{\textbf{Sensitivity of Huawei device-transfer results to input
orientation.} Results were calculated from median repeated out-of-fold
predictions for 103 paired physical samples. ``Orientation matched'' denotes
Huawei pixels rotated $90^\circ$ counterclockwise before the otherwise
unchanged preprocessing pipeline. Bias is prediction minus reference NTU.
FN and FP denote false-negative and false-positive errors, respectively,
at the 5-NTU operational threshold.}
\label{tab:supp_orientation}
\scriptsize
\setlength{\tabcolsep}{3.5pt}
\begin{tabular}{@{}llrrrrrrr@{}}
\toprule
Representation & Huawei input &
MAE & RMSE & Bias & $R^2$ & Accuracy & FN & FP \\
\midrule
Full image
 & Original stored pixels
 & 6.978 & 11.522 & $-6.755$ & 0.392 & 0.971 & 2 & 1 \\
Full image
 & Orientation matched
 & 6.457 & 10.373 & $-6.402$ & 0.507 & 0.981 & 2 & 0 \\
Target crop
 & Original stored pixels
 & 10.137 & 14.247 & $-10.131$ & 0.070 & 0.738 & 27 & 0 \\
Target crop
 & Orientation matched
 & 9.941 & 14.177 & $-9.934$ & 0.080 & 0.728 & 28 & 0 \\
\bottomrule
\end{tabular}
\end{table}
\FloatBarrier